\documentclass[]{aa}  

\usepackage{graphicx}
\usepackage{txfonts}
\usepackage[colorlinks = true,
            linkcolor = blue,
            urlcolor  = blue,
            citecolor = blue
            ]{hyperref}
\usepackage{amsmath}	% Advanced maths commands
\usepackage{xcolor}

\begin{document} 

% Variations in kilonova spectra from dynamical ejecta
   \title{Exploring the diversity of kilonovae with 3D radiative transfer}

   \subtitle{I. The polar direction}

   \author{C. E. Collins
          \inst{1}
          \and
          L. J. Shingles\inst{2} \and
          V. Vijayan\inst{3} \and
          A. Flörs\inst{3} \and
          O. Just\inst{3} \and
          F. McNeill\inst{2} \and
          Z. Xiong\inst{3} \and \\
          A. Bauswein\inst{3} \and
          K. Maguire\inst{1} \and
          S. A. Sim\inst{2}
          \fnmsep
          }
   \institute{School of Physics, Trinity College Dublin, The University of Dublin, Dublin 2, Ireland \\
              \email{ccollins@tcd.ie}
         \and
         Astrophysics Research Centre, School of Mathematics and Physics, Queens University Belfast, Belfast BT7 1NN, UK
        \and
        GSI Helmholtzzentrum f\"{u}r Schwerionenforschung, Planckstraße 1, 64291 Darmstadt, Germany
             % \thanks{The university of heaven temporarily does not
                     % accept e-mails}
             }

   \date{Received ; accepted }

\abstract{
We present 3D kilonova radiative transfer simulations for a series of binary neutron star merger models. The masses of the neutron stars are varied as well as the total mass of the system and two different equations of state were used (SFHO and DD2), producing a range in dynamical ejecta masses and elemental abundance patterns.
In this paper, we focus on the bolometric light curves and spectra in the polar direction for comparison with observations of the kilonova AT2017gfo.
We calculate line-by-line opacities and include new calibrated lanthanide atomic data.
All of the simulated spectra
show strong features from \ion{Sr}{II}, \ion{La}{III}, \ion{Gd}{III} and \ion{Ce}{III}, which appear to correspond to features identified in AT2017gfo, although the simulated features are generally more blueshifted.
The models with the lowest lanthanide fraction in the polar direction also show a \ion{Y}{II} feature.
\ion{Ce}{III}, \ion{Ce}{II}, \ion{Nd}{III} and \ion{Nd}{II} play an important role in shaping the spectral continuum.
While the bolometric luminosities in the polar direction vary with the ejecta mass of each model, we find only little sensitivity of the spectral properties to the merger configuration.
Our study demonstrates that dynamical ejecta alone can reproduce (although at earlier times) many spectral properties of AT2017gfo, suggesting dynamical ejecta may have a strong impact on the early spectral evolution. 
However, future simulations are needed to also elucidate the role of other ejecta components for shaping the kilonova spectrum.
}


   \maketitle
   \nolinenumbers
%
%-------------------------------------------------------------------

\section{Introduction}
In 2017, the kilonova AT2017gfo (e.g., \citealt{coulter2017a,cowperthwaite2017a,drout2017a,smartt2017a,villar2017a}, see \citealt{margutti2021a} for a review) was detected following the gravitational wave signal GW170817 \citep{abbott2017a}, resulting from the merging of binary neutron stars.
Since this detection there have been numerous efforts to interpret the observations of AT2017gfo.
A number of spectral features have been tentatively identified in the spectra of AT2017gfo, 
including \ion{Sr}{II} \citep{watson2019a}, \ion{La}{III} and \ion{Ce}{III} \citep{domoto2022a}, \ion{Y}{II} \citep{sneppen2023d} and \ion{Gd}{III} \citep{gillanders2024a, rahmouni2025a}. 
\citet{domoto2025a} have also shown that \ion{Th}{III} may be detectable. 
Additionally, \ion{Te}{III} has been suggested as a spectral feature in the nebular spectra of AT2017gfo \citep{hotokezaka2023a, gillanders2024a} as well as AT2023vfi (associated with GRB230307A \citealt{levan2024a, gillanders2025a}).
\citet{pognan2025a} and \citet{jerkstrand2026a} have suggested some additional features could be detectable in the nebular phase.
It has been suggested that the feature identified as \ion{Sr}{II} could be \ion{He}{I} \citep{tarumi2023a, perego2022a}, although \citet{sneppen2024c} argue that the rapid emergence of the feature is inconsistent with \ion{He}{I} and they favour the \ion{Sr}{II} interpretation, at least at the early epochs.
\citet{arya2026a} and \citet{chiba2026a} also discuss the contribution of \ion{He}{I} to this feature.

The atomic data for r-process elements still remains incompletely known. 
\citet{tanaka2020a} provide theoretical atomic data for r-process elements, which contains many energy levels and transitions required to simulate the total opacity in kilonovae. However, these theoretically calculated data are uncertain and not calibrated to experimentally known values. As a result, the wavelengths and strengths of the transitions are not accurate enough to predict individual spectral features.
This has been discussed by \citet{tanaka2020a} and \citet{domoto2021a}, and the impact on the \ion{Sr}{II} feature has been described by \citet{shingles2023a}.
Recently, \citet{floers2026a} have provided calibrated atomic data for singly and doubly ionized lanthanides.
In this paper, we make use of the atomic data from \citet{tanaka2020a}, but replace them where available with data from \citet{floers2026a}.
Most kilonova simulations have assumed local thermodynamic equilibrium (LTE), since the atomic data required to carry out non-LTE simulations are even more incomplete. 
LTE may be a reasonable assumption for the early phases of the kilonova, although \citet{brethauer2025a} have shown that non-thermal ionisation is important already during the first few days of the kilonova.
Some studies have carried out 1D non-LTE simulations for the nebular phase where LTE is no longer a valid assumption \citep{pognan2023a, pognan2025a, jerkstrand2026a}.

The aim of this work is to connect kilonova signatures with ejecta properties 
(e.g. mass and composition) %geometry I'll come to in the next paper with different directions
and with merger parameters (e.g. binary masses and the still unknown nuclear equation of state).
We carry out these simulations self-consistently by basing our radiative transfer on hydrodynamical simulations of neutron-star mergers. 
Detailed radiative transfer kilonova calculations performed on multidimensional simulation models are still rare in the literature \citep[e.g.][]{tanaka2013a, kasen2015a, miller2019a, kawaguchi2021a, kawaguchi2022a, shingles2023a},
in part due to the complexity of including the necessary physics ingredients (e.g. neutrino transport and small-scale turbulence) in hydrodynamic simulations and running them long enough to capture secular ejecta launched from the merger remnant. For this reason, many studies focus on the early ejecta component, called dynamical ejecta 
\citep[e.g.][]{bauswein2013a, wanajo2014a, radice2018a, just2022a, foucart2023a, neuweiler2023a, combi2023a, magistrelli2024a, rosswog2025a},
which tend to be relatively fast and may therefore surround a large fraction of the post-merger ejecta.
In AT2017gfo, it is likely that the dynamical ejecta component was not the dominant one in terms of mass \citep{siegel2019a, nedora2021a, just2023a, shingles2023a}. 

In this paper, we follow on from the work of \citet{shingles2023a},
who carried out 3D kilonova radiative transfer simulations based on dynamical ejecta from a single hydrodynamic neutron star merger model.
They calculated spectra using line-by-line opacities, and found that the spectra in the polar direction (near the orientation from which AT2017gfo was observed; \citealt{abbott2017b, perego2017a, dhawan2020a, mooley2022a}) show a good match to the spectra observed for AT2017gfo, although at earlier times than observed.
\citet{shingles2023a} demonstrated the importance of 3D radiative transfer simulations for kilonovae, showing that a 1D model did not reproduce the synthetic observables in any direction of the 3D model.
In this work, we expand on \citet{shingles2023a} to simulate the kilonova arising from a range of neutron star merger simulations with different neutron star masses and assuming different equations of state.
This allows us to explore the sensitivity of the light curves and spectral properties to different merger configurations and to make predictions for the variation that may be observed in future kilonovae.
Here we focus only on the spectra simulated from dynamical ejecta in the polar direction, near the observed viewing angle of AT2017gfo.

\begin{table*}
\centering
\small
\caption{Neutron star merger simulation parameters. For each model we give the mass of the secondary and primary neutron stars (M$_2$ and M$_1$, respectively), the mass ratio (q) between the neutron stars, the total mass of the two neutron stars (M$_\mathrm{tot}$)
and the ejecta mass (M$_\mathrm{ejecta}$) mapped into the radiative transfer simulation.
The mass-weighted mean lanthanide fraction, $\langle X_{\rm LN}\rangle_{\mathrm{M}}$ and mass-weighted mean electron fraction $\langle \mathrm{Y}_\mathrm{e}\rangle_{\mathrm{M}}$ 
are listed for each model, averaged over the entire model, as well as for the polar line of sight, 
$\langle X_{\rm LN}\rangle_{\mathrm{M, los}}$ and $\langle \mathrm{Y}_\mathrm{e}\rangle_{\mathrm{M,los}}$,
respectively.
The polar region is taken to be all model grid cells whose mid-points lie within $37^\circ$ of the pole.
Each of these were calculated from the ejecta models after mapping to the radiative transfer grid (rather than directly from the SPH particles).
Also shown are parameters for GW170817 \citep{abbott2017a} for comparison (the values inferred using low-spin priors) and the ejecta mass inferred for AT2017gfo by \citet{smartt2017a}.
}
\begin{tabular}{lcccccccrrrr}
\hline
       Label         & M$_2$    & {M$_1$} & {q}      & M$_\mathrm{tot}$     & M$_\mathrm{ejecta}$  & EoS  & $\langle X_{\rm LN}\rangle_{\mathrm{M}}$ & $\langle X_{\rm LN}\rangle_{\mathrm{M, los}}$ & $\langle \mathrm{Y}_\mathrm{e}\rangle_{\mathrm{M}}$ & $\langle \mathrm{Y}_\mathrm{e}\rangle_{\mathrm{M, los}}$\\ 
                                & {[M$_\odot$]}   & {[M$_\odot$]}& {}        & [M$_\odot$] & {[$10^{-2}$ M$_\odot$]}  &    &   &  &  &     \\ \hline  
       SFHO 1.3--1.3            & 1.3             & 1.3        & 1           & 2.6      &  0.353   & SFHO  &  0.021   &  0.0032  &  0.31  &  0.39 \\
       SFHO 1.35--1.35          & 1.35            & 1.35       & 1           & 2.7      &  0.487   & SFHO  &  0.022   &  0.0061  &  0.30  &  0.38   \\
       SFHO 1.375--1.375        & 1.375           & 1.375      & 1           & 2.75     &  0.584   & SFHO  &  0.021   &  0.0069  &  0.30  &  0.37 \\
       SFHO 1.4--1.4            & 1.4             & 1.4        & 1           &  2.8     &  0.919   & SFHO  &  0.019   &  0.0094  &  0.30  &  0.34   \\
       SFHO 1.11428--1.48571    & 1.11428         & 1.48571    & 0.75        & 2.6      &  0.988   & SFHO  &  0.024   &  0.011   &  0.27  &  0.34  \\
       SFHO 1.15714--1.54286    & 1.15714         & 1.54286    & 0.75        & 2.7      &  1.78    & SFHO  &  0.023   &  0.012   &  0.26  &  0.34 \\
       SFHO 1.17857--1.57143    & 1.17857         & 1.57143    & 0.75        & 2.75     &  1.42    & SFHO  &  0.021   &  0.011   &  0.28  &  0.34 \\
       SFHO 1.2--1.6            & 1.2             & 1.6        & 0.75        & 2.8      &  1.41    & SFHO  &  0.021   &  0.0075  &  0.28  &  0.35  \\
       DD2 1.3--1.3             &  1.3             &   1.3     &   1         & 2.6      &  0.348   & DD2   &  0.017   &  0.0048  &  0.33  &  0.42 \\
       DD2 1.35--1.35           & 1.35            & 1.35       & 1           & 2.7      &  0.382   & DD2   &  0.019   &  0.0071  &  0.31  &  0.38 \\
       DD2 1.375--1.375         & 1.375           & 1.375      & 1           & 2.75     &  0.347   & DD2   &  0.018   &  0.0015  &  0.31  &  0.40 \\
       DD2 1.4--1.4             & 1.4             & 1.4        & 1           &  2.8     &  0.401   & DD2   &  0.020   &  0.0064  &  0.30  &  0.37  \\
       DD2 1.45--1.45           & 1.45             & 1.45      & 1           &  2.9     &  0.443   & DD2   &  0.022   &  0.0061  &  0.30  &  0.37  \\
       DD2 1.11428--1.48571     & 1.11428         & 1.48571    & 0.75        & 2.6      &  0.830   & DD2   &  0.033   &  0.0091  &  0.24  &  0.35 \\
       DD2 1.15714--1.54286     & 1.15714         & 1.54286    & 0.75        & 2.7      &  0.915   & DD2   &  0.030   &  0.017   &  0.24  &  0.31 \\
       DD2 1.17857--1.57143     & 1.17857         & 1.57143    & 0.75        & 2.75     &  0.868   & DD2   &  0.031   &  0.011   &  0.24  &  0.34 \\
       DD2 1.2--1.6             & 1.2             & 1.6        & 0.75        & 2.8      &  1.03    & DD2   &  0.029   &  0.016   &  0.25  &  0.32 \\
       DD2 1.24286--1.65714     & 1.24286          & 1.65714    & 0.75       & 2.9      &  1.09    & DD2   &  0.025   &  0.015   &  0.26  &  0.31   \\
                    \hline \hline
GW170817/AT2017gfo  &    1.17--1.36  &  1.36--1.60 &  0.7--1.0  & 2.74${^{+0.04}_{-0.01}}$  &  4.0$\pm$1.0 & & & \\ \hline
\label{tab:models}
\end{tabular}
\end{table*}

\section{Methods}

\subsection{Radiative transfer}

We carry out 3D radiative transfer simulations using the time-dependent Monte Carlo radiative transfer code \textsc{artis} \citep{sim2007b, kromer2009a, bulla2015a, shingles2020a, shingles2023a} (based on the methods of \citealt{lucy2002a, lucy2003a, lucy2005a}), in the same way as \citet{shingles2023a}.
Radioactive decays are followed time-dependently by coupling a nuclear decay network to \textsc{artis}.
This includes a Monte Carlo $\gamma$-ray transport scheme \citep{sim2007b} and a time-dependent treatment of $\alpha$ and $\beta^-$ particle thermalisation (see \citealt{shingles2023a} for details).
\citet{shingles2023a} showed that the nuclear decay network coupled to \textsc{artis} is in good agreement with the full network calculation.
Line-by-line opacities are calculated using the Sobolev approximation. This line-by-line approach enables a treatment of fluorescence and also allows the direct association of spectral features with the atomic transitions.
We assume local thermodynamic equilibrium (LTE).

Each simulation propagates $1.44 \times 10^8$ Monte Carlo packets and is carried out between 0.1 and 80 days, with 64 logarithmically spaced time steps.
Escaping packets of radiation are binned into 100 uniformly spaced angle bins in cos($\theta$) and $\phi$ directions to generate angle-dependent light curves and spectra.
In this paper we focus on the synthetic observables in the region $\theta < 37 ^\circ$ (averaging over all $\phi$ angles),
which is the region close to the pole (chosen to be the $-z$ axis).
This averaging reduces the Monte Carlo noise from the simulation.
North to south asymmetries are present in the models of $\sim10$s of percent, as previously discussed by \citet{collins2024a}. This leads to quantitive differences at opposite poles, but qualitatively similar spectra.

\subsubsection{Atomic data}

As in \citet{shingles2023a}, we source atomic data for low mass elements from the CMFGEN compilation\footnote{\tiny \url{http://kookaburra.phyast.pitt.edu/hillier/web/CMFGEN.htm}} \citep{hillier1990a} (up to $Z=28$). Most of the rest of our atomic data is sourced from the Japan-Lithuania Opacity Database for Kilonova\footnote{\href{http://dpc.nifs.ac.jp/DB/Opacity-Database/}{http://dpc.nifs.ac.jp/DB/Opacity-Database/} (Version 2.1)} \citep{tanaka2020a}. 
We replace lanthanide atomic data with calibrated data from \citet{floers2026a}\footnote{\href{https://zenodo.org/records/15835361}{https://zenodo.org/records/15835361}}.
\citet{floers2026a} have calibrated theoretical energy levels to experimentally known values where possible. This includes \ion{La}{II--III},
\ion{Ce}{II--III},
\ion{Pr}{II--III},
\ion{Nd}{II--III},
\ion{Pm}{II},
\ion{Sm}{II--III},
\mbox{\ion{Eu}{II--III}},
\ion{Gd}{II--III},
\ion{Tb}{II--III},
\ion{Dy}{II},
\ion{Ho}{II--III},
\ion{Er}{II--III},
\ion{Tm}{II--III} and
\ion{Yb}{II--III}.
We also replace \ion{Sr}{I--V}, \ion{Y}{I--II}, \ion{Zr}{I--IV} and \ion{Ba}{I--II} with atomic data from Kurucz \citep{kurucz2018a, kurucz2006a}, as well as neutral species where available, including \ion{La}{I}, \ion{Ce}{I}, \ion{Pr}{I}, \ion{Nd}{I}, \ion{Sm}{I}, \ion{Eu}{I}, \ion{Gd}{I}, \ion{Dy}{I}, \ion{Er}{I}, \ion{Tm}{I}, \ion{Yb}{I}, \ion{Lu}{I}, \ion{Hf}{I}, \ion{Ta}{I}, \ion{W}{I}, \ion{Re}{I}, \ion{Os}{I}, \ion{Ir}{I}, \ion{Pt}{I}, \ion{Au}{I}, \ion{Hg}{I}, \ion{Pb}{I} and \ion{Bi}{I}.

\subsection{Neutron star merger simulations}

The hydrodynamical neutron star merger simulations we consider were presented by \citealt{vijayanthesis} and Vijayan et al. (in prep.; see also \citealt{vijayan2026a}). 
These were simulated using a 3D general relativistic smoothed-particle hydrodynamics (SPH) code \citep{oechslin2002a, bauswein2010a} and included a leakage-plus-absorption neutrino treatment (ILEAS \citealt{ardevol2019a}).
The key properties of the models are summarised in Table~\ref{tab:models}.
The suite of models tests varying neutron star masses with different binary mass ratios and different total masses.
Two equations of state are employed; SFHO \citep{steiner2013a} and DD2 \citep{typel2010a}. SFHO is a relatively soft equation of state, while DD2 is a relatively stiff equation of state.
SFHO typically leads to higher ejecta masses than DD2 \citep{bauswein2013a, sekiguchi2015a, radice2018a, kullmann2022a}.
The inferred properties for GW170817/AT2017gfo are also shown for comparison.

The merger simulations were carried out for 20 ms
after the stars merged, and therefore the simulations only capture dynamical ejecta. Mass ejection is expected to continue beyond the time these simulations ended (for a discussion of secular ejecta see, e.g. \citealt{fernandez2013a, perego2014a, just2015a, fujibayashi2018a, siegel2018a}). In future studies, we will investigate the impact of later ejecta components, however in this work we carry out radiative transfer simulations for the dynamical ejecta component only.
Owing to this, the model ejecta masses listed in Table~\ref{tab:models} are all relatively low, and smaller than the mass inferred from AT2017gfo.

The merger simulations were mapped to a $50^3$ Cartesian grid for the radiative transfer simulation using the same approach as in \citet{collins2023a}, except that we do not assume any further expansion of the SPH particles following the hydrodynamics simulation before mapping to the radiative transfer grid. Instead, the SPH particles are mapped at their final position and we assume the ejecta expand homologously (assuming $v = r/t$, where $v$ is velocity, $r$ is radius and $t$ is time since the merger) once mapped to the Cartesian grid. 
Note that this is an approximation 
since at 20\,ms a fair fraction of energy carried by the ejecta is still in the form of thermal energy and because late-time r-process heating is neglected (e.g. \citealt{foucart2021a, just2025a}). 
Both can induce moderate changes of the velocity structure, however, mainly in the slower ($v \lesssim 0.15$\,c) part of the ejecta, which is likely less relevant for the early kilonova signal studied here.

\subsection{Nucleosynthesis}

Nucleosynthesis calculations were carried out in the same way as described by \citet{collins2023a} for each SPH particle trajectory, using the same nuclear reaction network as in \citet{mendoza2015a} together with the set of nuclear reactions labelled `FRDM'.
The elemental abundance profiles are mapped onto the merger ejecta at 0.1 days and we initialise Monte Carlo energy packets with the energy released up to this time as calculated from the nuclear network (accounting for energy lost to the expansion of the ejecta). 
After 0.1 days, energy deposition is calculated from the radioactive decays by \textsc{artis}. 

\begin{figure*}
    \centering
    \includegraphics[width=0.9\linewidth]{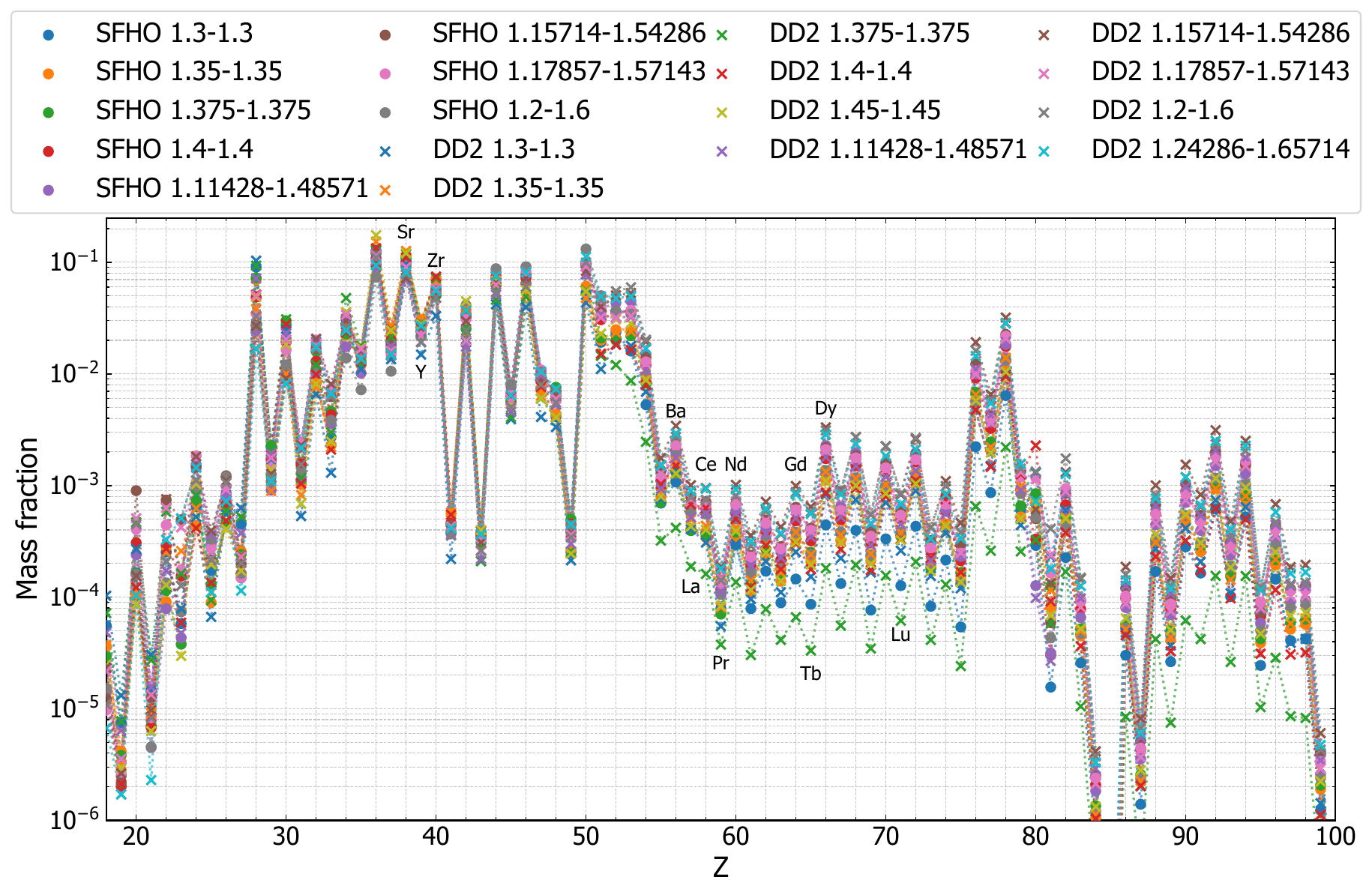}
    \caption{Elemental mass fractions
    in the ejecta in the region of the pole for each model at 0.1 days. The polar region is defined as all model grid cells whose mid-points lie within the region around the pole ($<37^\circ$). 
    Circle markers indicate models using the SFHO equation of state while x markers show models using the DD2 equation of state.
    }
    \label{fig:abundances}
\end{figure*}

In Figure~\ref{fig:abundances} we show the mass fractions in the polar region (defined as the region within $37^\circ$ around the polar axis) for each element synthesised in each of the merger models.
In the polar region, the greatest difference between the models is in the lanthanide mass fractions, where DD2 1.375--1.375 and SFHO 1.3--1.3 show relatively low abundances of these elements.
The $Y_\mathrm{e}$ distribution for the models with the highest and lowest lanthanide fraction in the polar region are shown in Figure~\ref{fig:Ye_distribution} in Appendix~\ref{appendix}.

The entire ejecta show on average very similar abundance patterns between each of the models (see Figure~\ref{fig:mass_fractions_entire_ejecta} in Appendix~\ref{appendix}), with relatively small differences across all models. The spherically averaged, mass weighted lanthanide fractions and $Y_\mathrm{e}$ for these models are listed in Table~\ref{tab:models}.
The ejecta near the poles tend to have lower lanthanide fractions since these directions are irradiated by neutrinos, increasing the $Y_\mathrm{e}$ of the material. The ejecta in the plane of the merger tend to have higher lanthanide fractions and a lower $Y_\mathrm{e}$, and also tends to have more mass ejected in these directions.
The DD2 models show a larger variation in the spherically averaged, mass weighted lanthanide fractions and $Y_\mathrm{e}$ than the SFHO models.

\begin{figure*}
    \centering
    \includegraphics[width=0.98\linewidth]{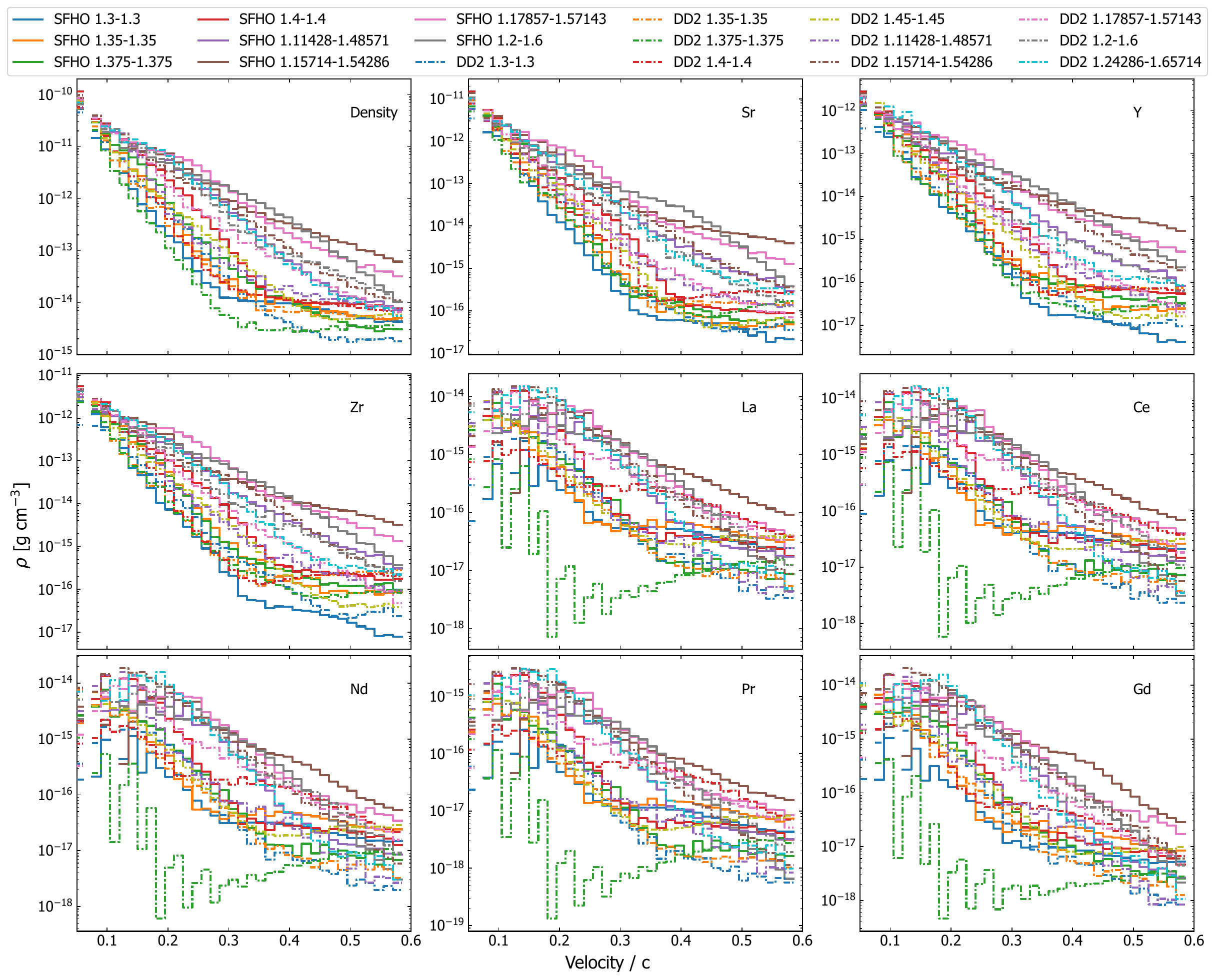}
    \caption{Model density and density of select elements in the region near the pole at 0.1 days.
    Model grid cells with mid-points lying between the pole and $37^\circ$ have been binned by radial velocity and the average density within each velocity bin is plotted.
    }
    \label{fig:model_los}
\end{figure*}

We also show the mean density in this region in Figure~\ref{fig:model_los}, as well as the density of select elements.
When the ejecta are optically thick a significant fraction of the radiation reaching an observer near the poles should come from ejecta in the polar line of sight, however, we caution that as the ejecta become optically thin it is expected that radiation would be able to travel further across the ejecta and radiation reaching an observer would come from broader ranges of ejecta than just this line of sight (as has been discussed by \citealt{collins2024a}).

\begin{figure}
    \centering
    \includegraphics[width=\linewidth]{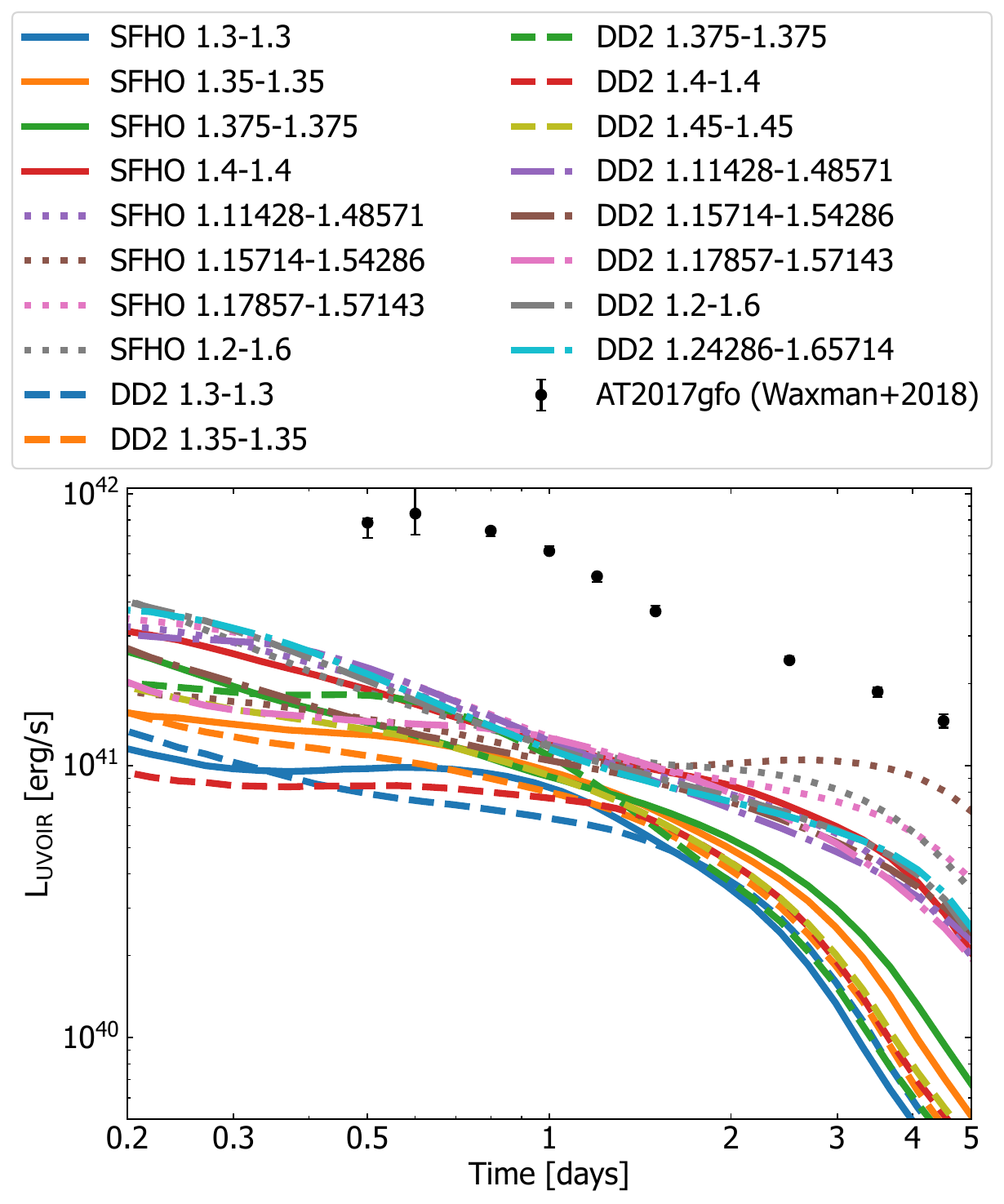}
    
    \caption{Bolometric light curves for each of the merger models at the pole. Solid and chain-dashed lines show models with equal binary neutron star masses, using the SFHO or DD2 equation of state, respectively. The dotted and dash-dotted lines show models using SFHO or DD2 with unequal neutron star masses (with a mass ratio of 0.75). Models with the same initial neutron star masses but different equation of state have the same colour. 
    }
    \label{fig:bollightcurves}
\end{figure}

\section{Results}

We discuss the bolometric light curves and spectra predicted for the models in the polar direction.
This includes the estimated direction from which AT2017gfo was observed of around 20 -- 30$^{\circ}$ from the pole \citep{abbott2017b,perego2017a, dhawan2020a, mooley2022a}. 
% abbott2017b -- <36deg or <28 deg depending on H0
% perego2017a -- between 15 and 35 deg
% dhawan2020a 32.5 deg
% mooley2022a 19-25 deg
We compare these simulated observables to the observations of AT2017gfo.

\subsection{Bolometric Light curves}

The bolometric light curves in the polar directions are shown for each of the merger models in Figure~\ref{fig:bollightcurves}.
For ease of comparison, we show the light curves at one pole (calculated by binning all escaping Monte Carlo packets of radiation escaping in the directions $<37^\circ$ from the pole, in all azimuthal directions).
As expected from the low ejecta masses (since only dynamical ejecta is included in these simulations), the model light curves are fainter than AT2017gfo and they fade faster than the observations. 

Similarly to \citet{collins2023a}, we find that all of the models decline in bolometric luminosity from very early times, and the peak of the bolometric light curve is before the start of the simulation (0.1 days). 
The simulated light curve brightness generally increases with increasing ejecta mass.
The light curve decline rate also slows down for more massive models, such that the more massive models remain brighter for longer.
As is shown in Table~\ref{tab:models}, the ejecta mass increases with increasing initial neutron star mass, but also increases for unequal mass binaries \citep[see e.g.][]{bauswein2013a}.
The models using the SFHO equation of state eject more mass than the models using DD2.
The most massive ejecta model is SFHO~1.15714--1.54286.
At times less than $\sim 1$ day this model is not the brightest, but it does become brightest at $\sim 1.1$ days, as it rises to a secondary peak. This suggests that relatively more energy is trapped for longer by the ejecta in this model at early times compared to other models.
The most massive models show this rise to a peak after $\sim 1$ day, while the less massive models show a declining light curve right from the start of the calculation (0.1 days) and exhibit a shoulder where the decline rate increases.
The shoulder is earlier for the less massive models and occurs later with increased ejecta mass.
This demonstrates a general trend where lower ejecta mass leads to a faster bolometric decline rate.

\subsection{Spectra}
\label{sec:spectra}

\begin{table}
    \centering
    \caption{Time where the SED of each simulation most closely matches the SED of AT2017gfo at four observed epochs.}
    \resizebox{\columnwidth}{!}{
    \begin{tabular}{lcccc}
\hline
Label & Epoch 1 & Epoch 2 & Epoch 3 & Epoch 4 \\
           & [days]  & [days]  & [days]  & [days] \\
\hline
SFHO 1.3-1.3 & 0.18 & 0.90 & 1.20 & 1.50 \\
SFHO 1.35-1.35 & 0.20 & 0.85 & 1.00 & 1.30 \\
SFHO 1.375-1.375 & 0.25 & 0.70 & 1.00 & 1.20 \\
SFHO 1.4-1.4 & 0.28 & 0.70 & 0.90 & 1.20 \\
SFHO 1.11428-1.48571 & 0.34 & 0.80 & 1.00 & 1.30 \\
SFHO 1.15714-1.54286 & 0.25 & 0.80 & 1.00 & 1.30 \\
SFHO 1.17857-1.57143 & 0.32 & 0.85 & 1.00 & 1.30 \\
SFHO 1.2-1.6 & 0.30 & 0.70 & 0.90 & 1.40 \\
DD2 1.3-1.3 & 0.25 & 0.75 & 0.90 & 1.10 \\
DD2 1.35-1.35 & 0.22 & 0.70 & 1.00 & 1.10 \\
DD2 1.375-1.375 & 0.36 & 1.20 & 1.40 & 1.50 \\
DD2 1.4-1.4 & 0.18 & 0.75 & 1.00 & 1.20 \\
DD2 1.45-1.45 & 0.22 & 0.75 & 1.00 & 1.20 \\
DD2 1.11428-1.48571 & 0.45 & 0.85 & 1.00 & 1.30 \\
DD2 1.15714-1.54286 & 0.25 & 0.80 & 1.00 & 1.30 \\
DD2 1.17857-1.57143 & 0.22 & 0.95 & 1.25 & 1.50 \\
DD2 1.2-1.6 & 0.40 & 0.80 & 1.00 & 1.20 \\
DD2 1.24286-1.65714 & 0.35 & 0.85 & 1.00 & 1.20 \\
\hline \hline
GW170817/AT2017gfo  & 1.43 & 2.42 & 3.41 & 4.40 \\
\hline
\end{tabular}
}
    \label{tab:epochtimes}
\end{table}

\begin{figure*}
    \centering
    \includegraphics[width=0.98\linewidth]{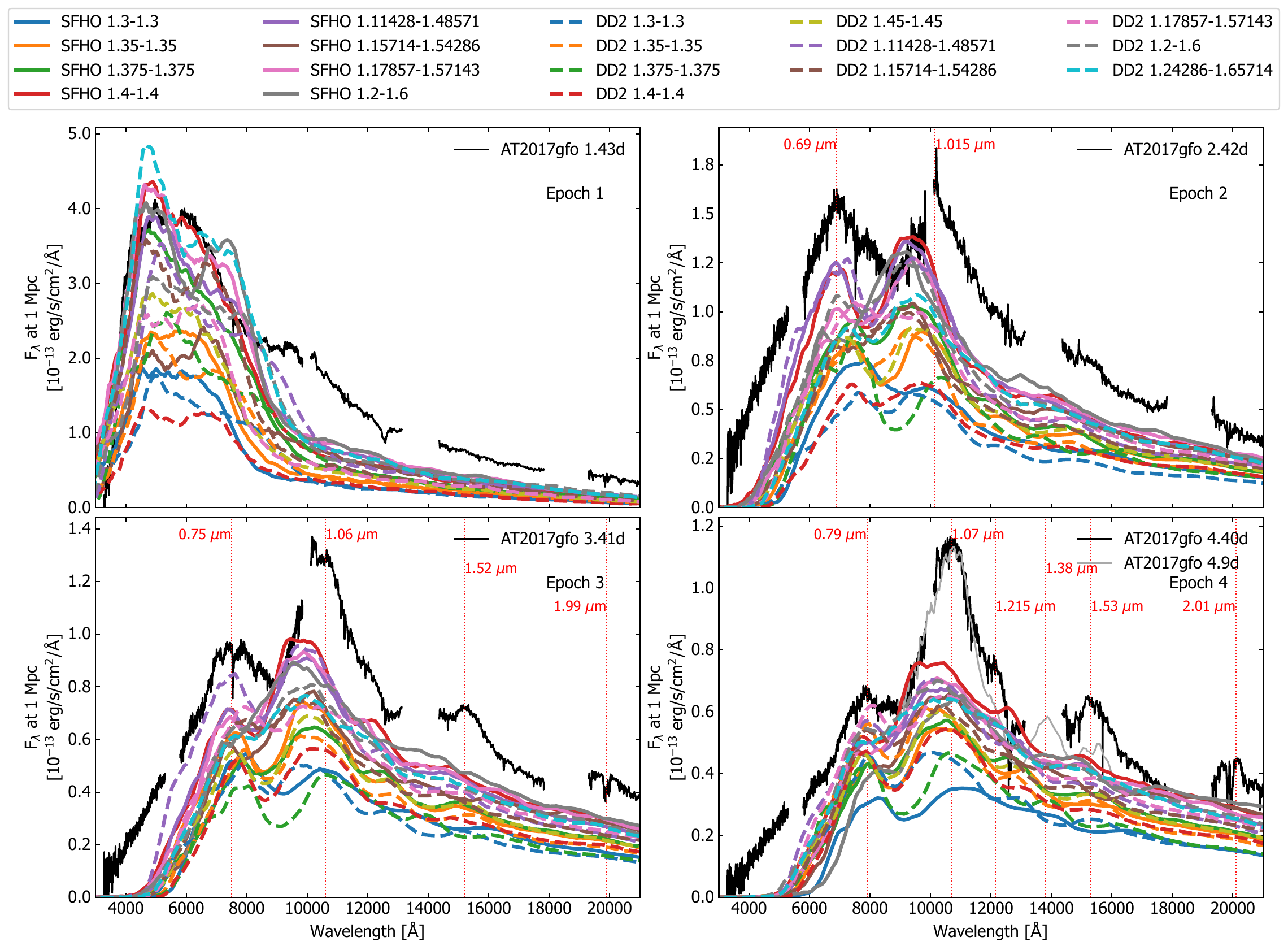}
    \caption{All model spectra at the pole compared to the observed spectra of AT2017gfo \citep{smartt2017a,pian2017a,tanvir2017a}. Note the difference in time between simulations and observations listed in Table~\ref{tab:epochtimes}.
    The vertical lines indicate the wavelengths of emission-like features that emerge in AT2017gfo (wavelengths taken from \citealt{gillanders2024a}).}
    \label{fig:spectra_poles}
\end{figure*}

\begin{figure*}
    \centering
    \includegraphics[width=\linewidth]{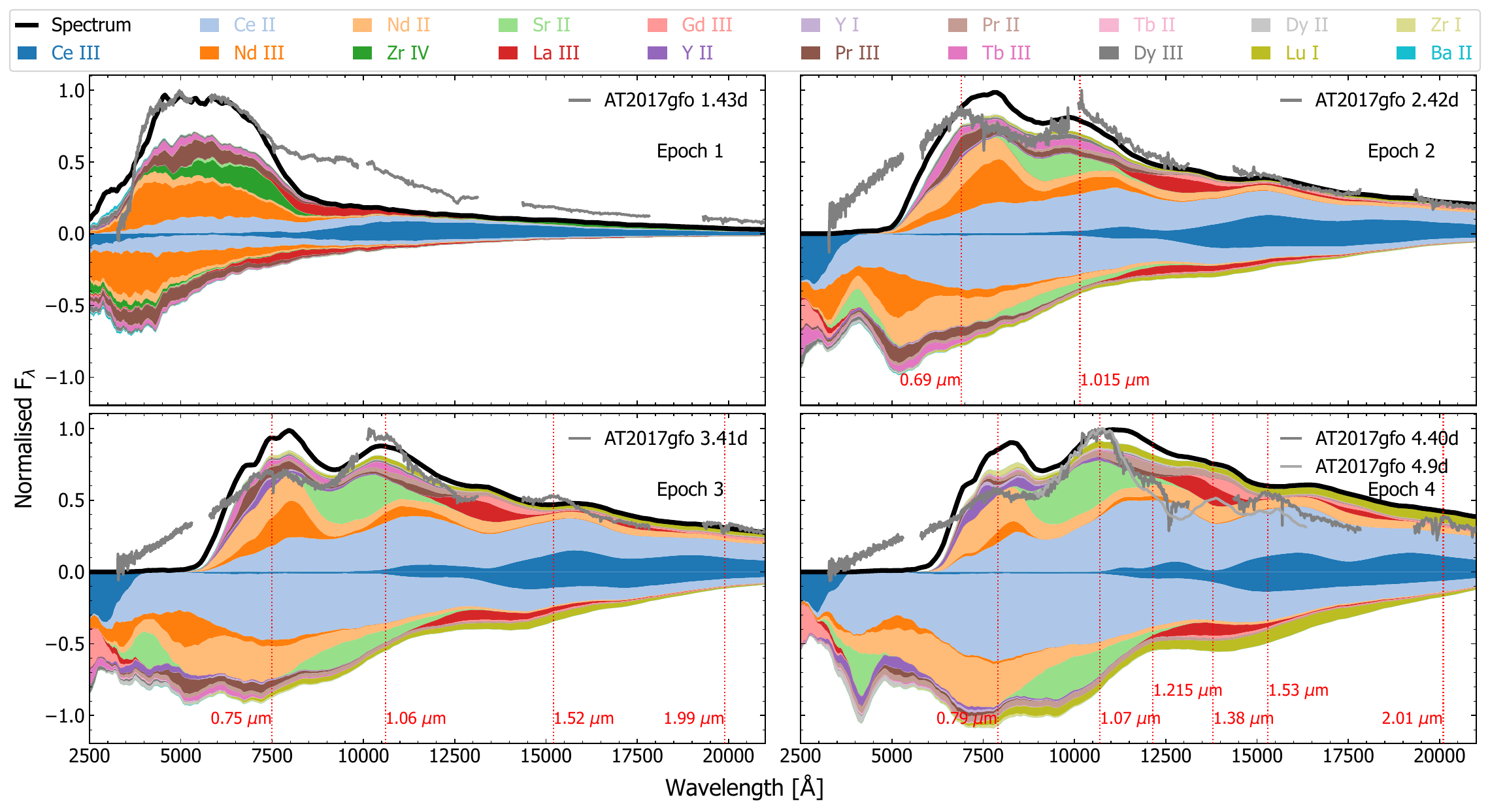}
    \includegraphics[width=0.95\linewidth]{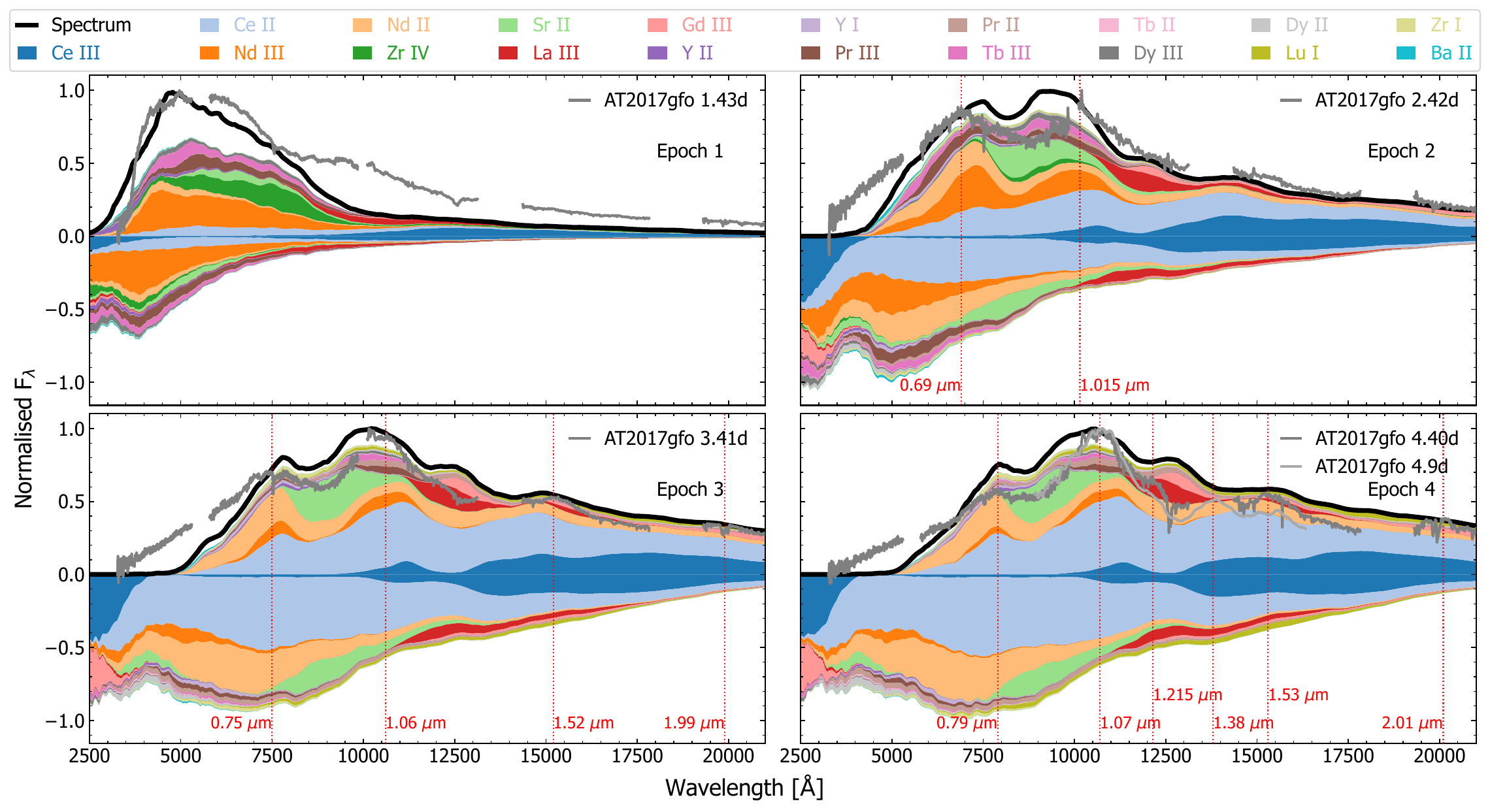}
    \caption{Spectra for SFHO 1.3--1.3 (upper) and SFHO 1.375--1.375 (lower) scaled and compared to AT2017gfo at epochs 1--4 (see Table~\ref{tab:epochtimes} for simulation times).
    The shaded regions indicate the key ion contributions to shaping the spectra. Above the x-axis are the contributions to emission while below the x-axis the main ion contributions to absorption are indicated.
    }
    \label{fig:specemissionabsorption}
\end{figure*}

Again, for ease of comparison, we discuss the spectra from one pole.
The spectra are compared to the observations of AT2017gfo in Figure~\ref{fig:spectra_poles}.
A Savitzky–Golay filter is applied to the simulated spectra to reduce the appearance of Monte Carlo noise.
We show the AT2017gfo observations by \citet{pian2017a} and \citet{smartt2017a}\footnote{Specifically we use the dereddened and deredshifted spectra provided by ENGRAVE at \href{http://www.engrave-eso.org/AT2017gfo-Data-Release/}{http://www.engrave-eso.org/AT2017gfo-Data-Release/}} and assume a distance of 40 Mpc \citep{smartt2017a}. 

The simulation times shown in Figure~\ref{fig:spectra_poles} have been chosen so that the spectral energy distribution (SED) of each simulated spectrum is the closest match to the SED of the observations of AT2017gfo at each epoch (assessed visually).
These times are listed in Table~\ref{tab:epochtimes}.
Note that the times where the SED of the simulations show the best match to the observations are much earlier than the observed times, and the evolution of the spectral series is faster than observed for AT2017gfo.
This is most likely due to the low ejecta mass in the models compared to the inferred mass of AT2017gfo, since we only consider dynamical ejecta here.
At the times where the models are compared to the observations, the brighter simulations are of comparable brightness to AT2017gfo (Figure~\ref{fig:spectra_poles}).

The models generally show similar trends, where initially the SED peaks at blue wavelengths and the spectra are somewhat featureless (see the epoch 1 spectra in Figure~\ref{fig:spectra_poles}), although note that some models have already developed a feature due to \ion{Sr}{II} at this first epoch (see Section~\ref{sec:epoch1} for further discussion).
Figure~\ref{fig:specemissionabsorption} shows SFHO~1.3--1.3 and SFHO~1.375--1.375 as representative models, and indicates the main ion contributions to emission and to absorption in these models, which are similar across all of the models (further discussion in Sections~\ref{sec:epoch1}--\ref{sec:epoch4}). 
In \textsc{artis}, the last interaction of a photon before it escapes is recorded, and we use this to colour code the relative emission and absorption contributions in Figure~\ref{fig:specemissionabsorption}.
The emission contributions are determined by recording the last interaction of a Monte Carlo packet of photons in \textsc{artis} before it escaped the ejecta. The absorption contributions are determined by the ion that last absorbed the packet before it went on to escape the ejecta.

From epoch 2 onward, most models develop a broad feature near the peak of the SED, as can be seen in Figure~\ref{fig:specemissionabsorption} for SFHO~1.3--1.3 and SFHO~1.375--1.375. This is similar to the feature observed in AT2017gfo. In the simulations this is primarily due to the \ion{Sr}{II} triplet, which forms a P-Cygni like feature near the peak of the SED, as was inferred for AT2017gfo \citep{watson2019a}. Over time, the peak of the SED evolves to redder wavelengths, which is consistent with the observations of AT2017gfo, and more spectral features develop, including a blended \ion{La}{III} and \ion{Gd}{III} feature and \ion{Ce}{III} features in addition to the \ion{Sr}{II} feature (discussed in more detail in Sections~\ref{sec:epoch1}--\ref{sec:epoch4}).
The simulated spectral features are broad due to the high ejecta velocities of the models.

\begin{figure*}
    \centering
    \includegraphics[width=0.48\linewidth]{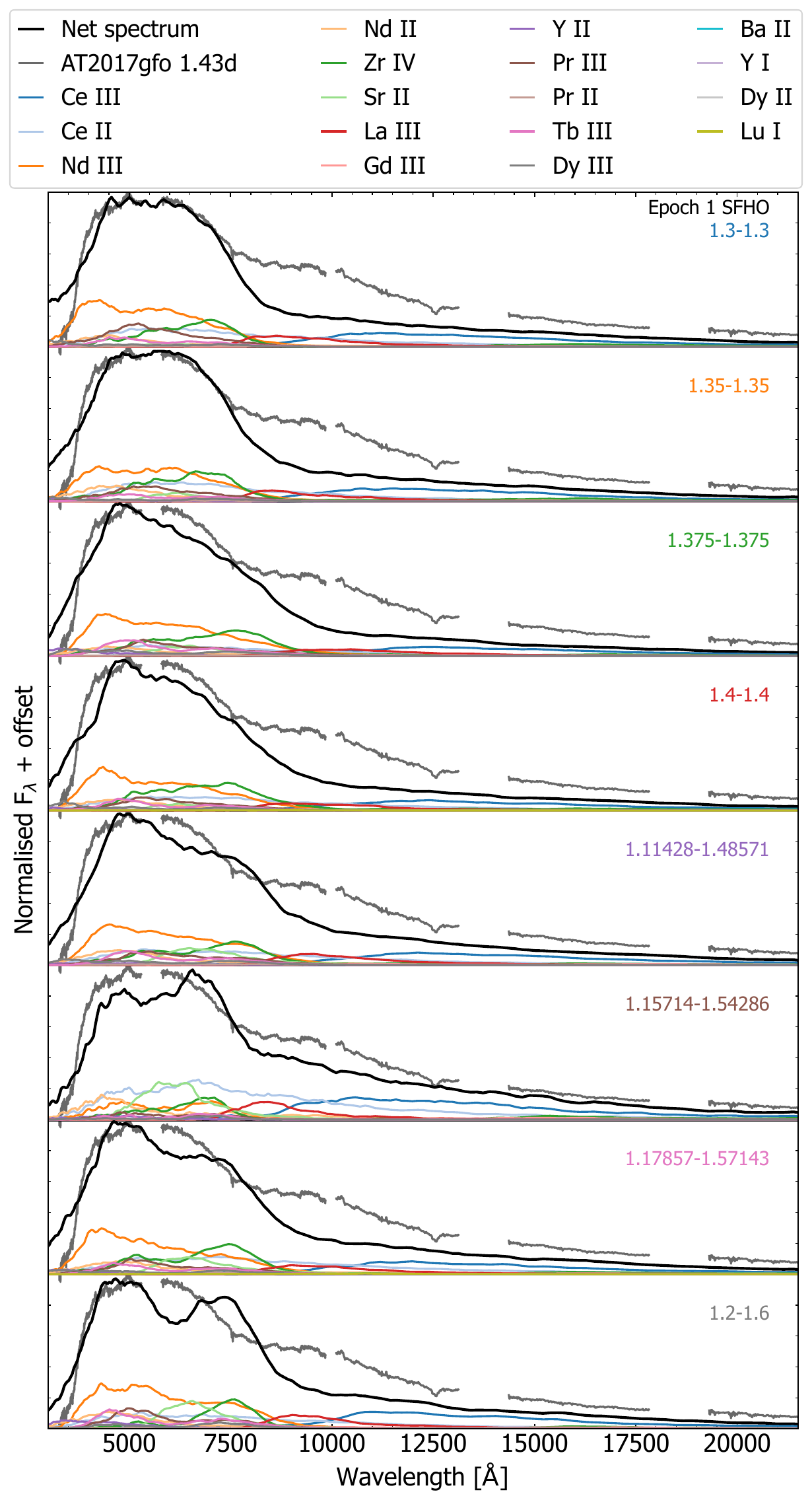}
    \includegraphics[width=0.48\linewidth]{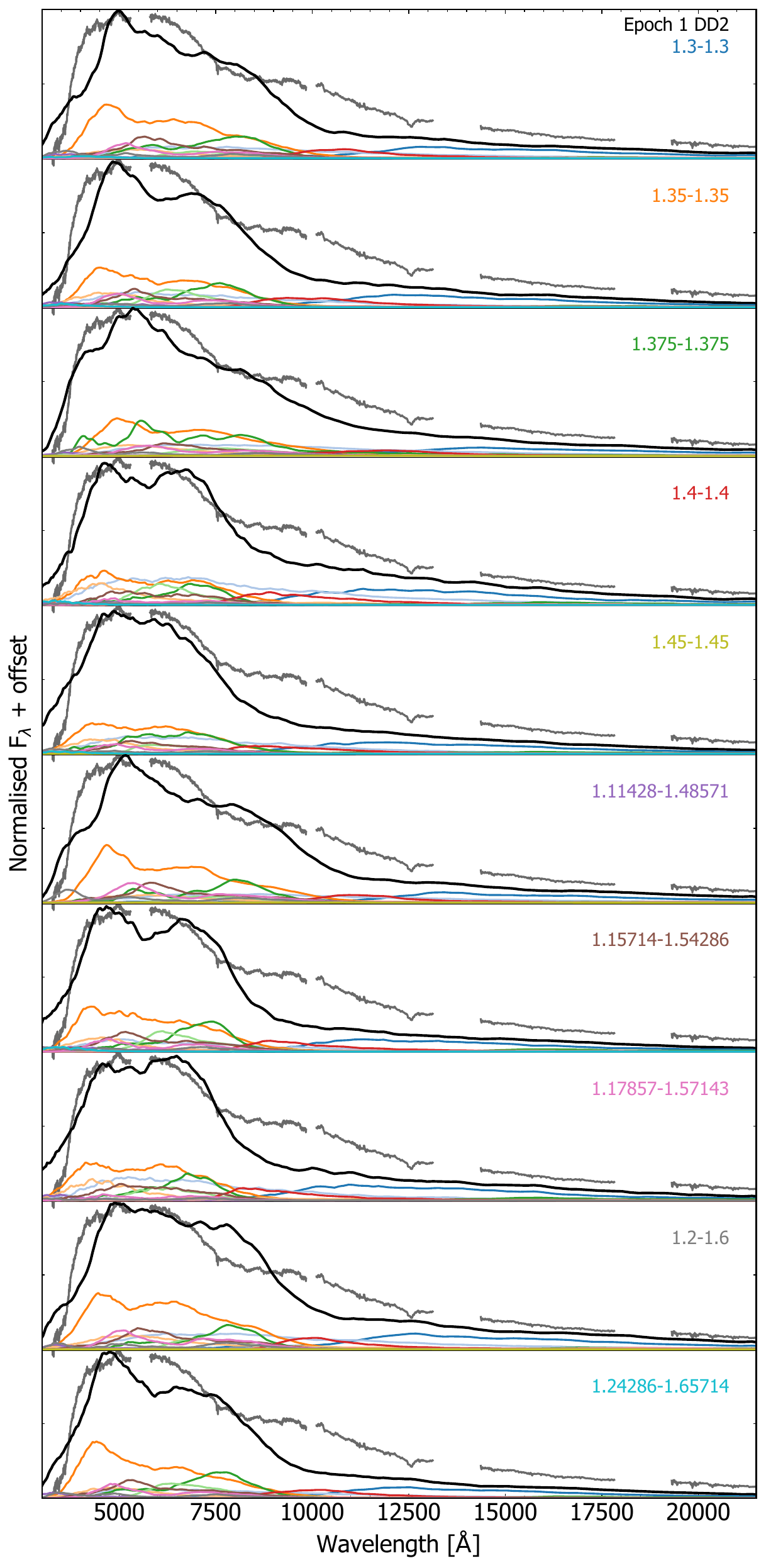}
    \caption{Scaled simulated spectra at epoch 1 (see times in Table~\ref{tab:epochtimes}) for all models compared to the observations of AT2017gfo at 1.43 days.
    The emission from select ions (i.e. ions that the photons last interacted with before escaping the  ejecta) is shown for each model (similarly to Figure~\ref{fig:specemissionabsorption}, except without stacking the ion contributions), indicating the most important contributions shaping the spectrum.
    }
    \label{fig:spec_emission_epoch1}
\end{figure*}

\begin{figure*}
    \centering
    \includegraphics[width=0.48\linewidth]{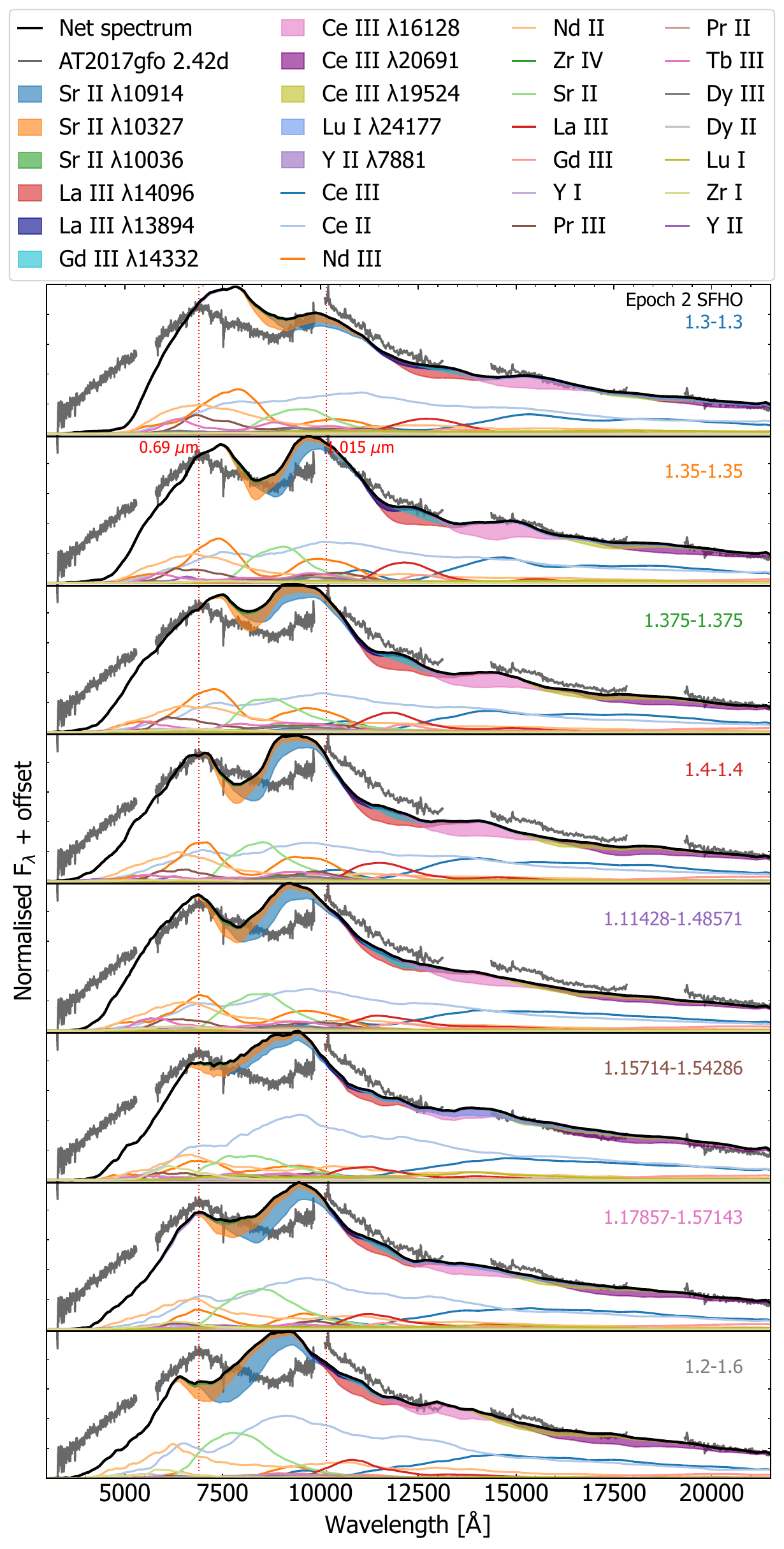}
    \includegraphics[width=0.48\linewidth]{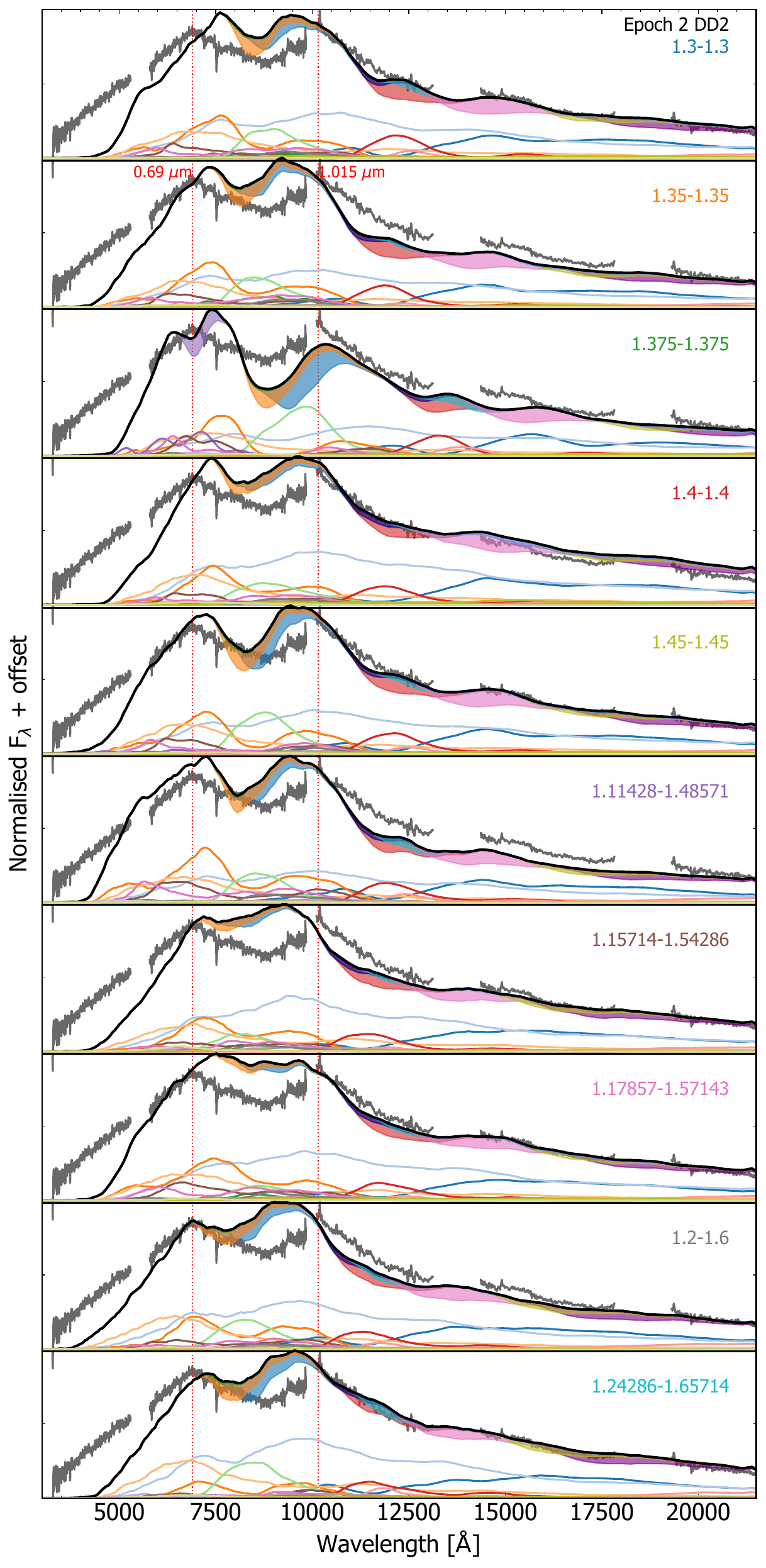}
    \caption{Same as Figure~\ref{fig:spec_emission_epoch1} for the spectra at epoch 2, where the ion emission contributions are shown beneath each model spectrum.
    In addition to the ion contributions (shown by solid coloured lines), we also show the fraction of the ion contribution made up of the strongest individual spectral lines. These are indicated by the shaded regions hanging down from each spectrum, where the shading corresponds to the strength of the emission contribution from each individual spectral line.
    This shows the spectral regions shaped by each of these individual line transitions.}
    \label{fig:spec_emission_epoch2}
\end{figure*}

\begin{figure*}
    \centering
    \includegraphics[width=0.48\linewidth]{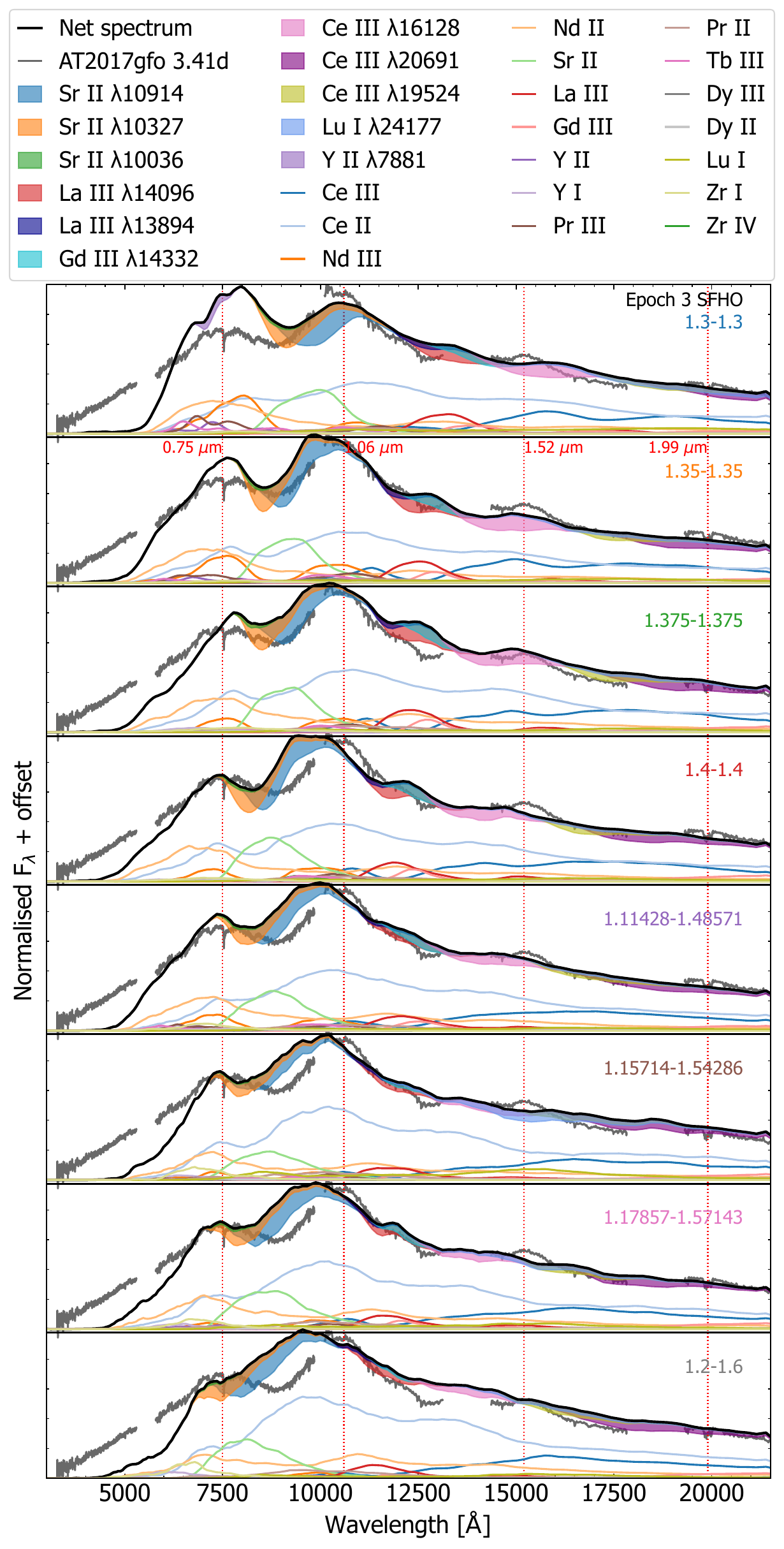}
    \includegraphics[width=0.48\linewidth]{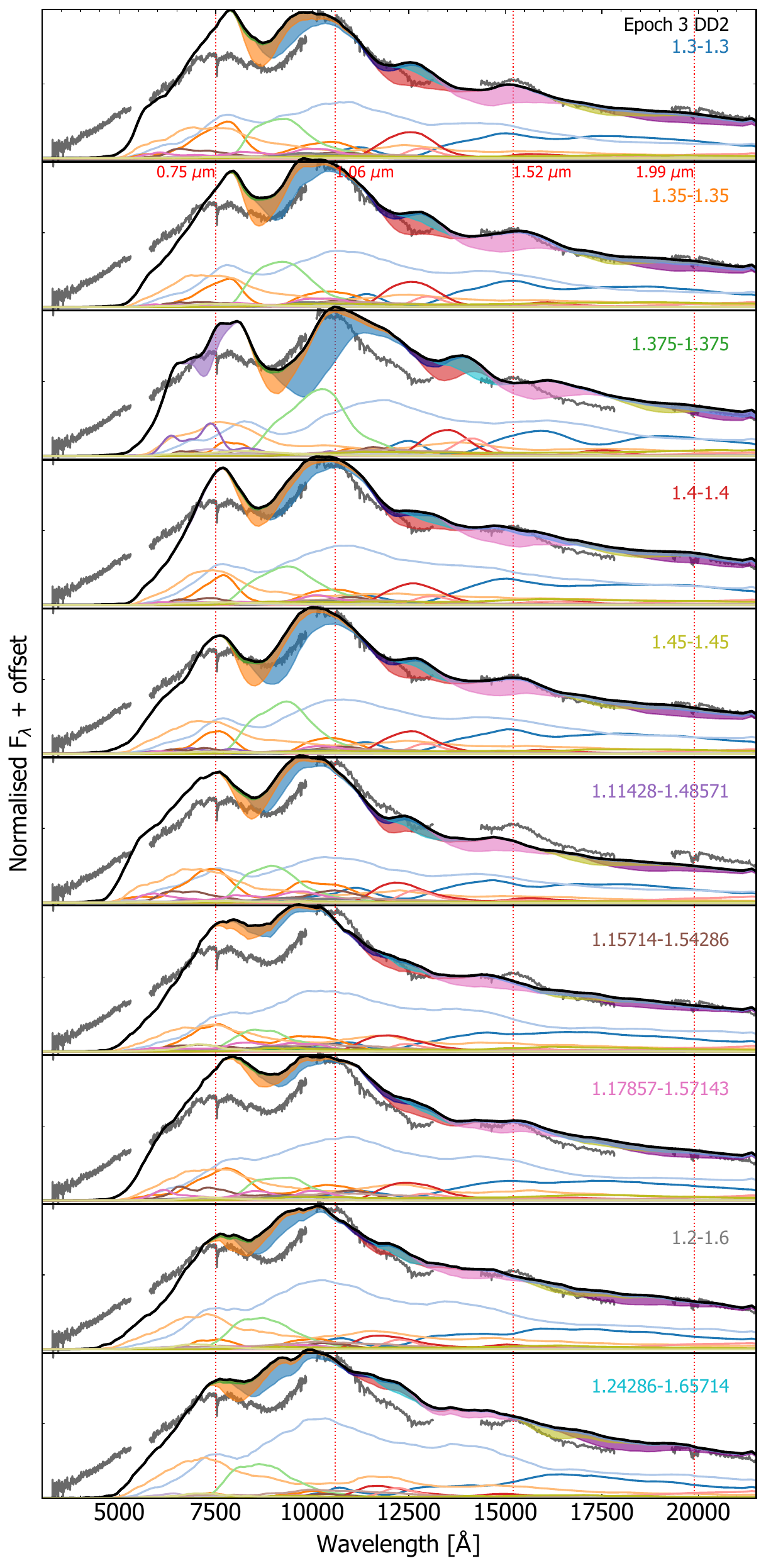}
    \caption{Same as Figure~\ref{fig:spec_emission_epoch2} for the spectra at epoch 3. Beneath each spectrum, the solid lines show the ion emission contribution while shaded regions show the fraction of the ion contribution emitted by individual spectral lines.}
    \label{fig:spec_emission_epoch3}
\end{figure*}

\begin{figure*}
    \centering
    \includegraphics[width=0.48\linewidth]{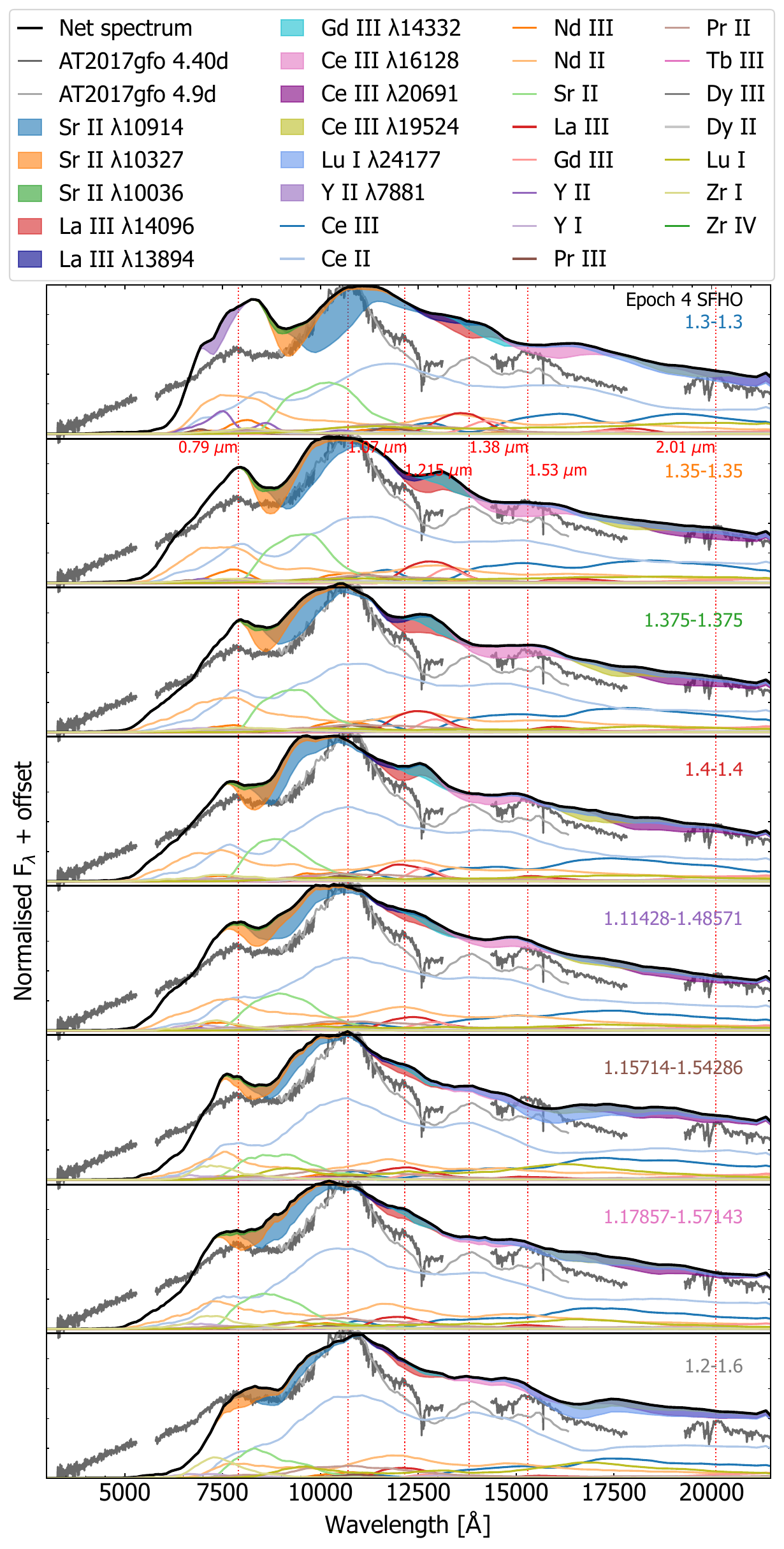}
    \includegraphics[width=0.48\linewidth]{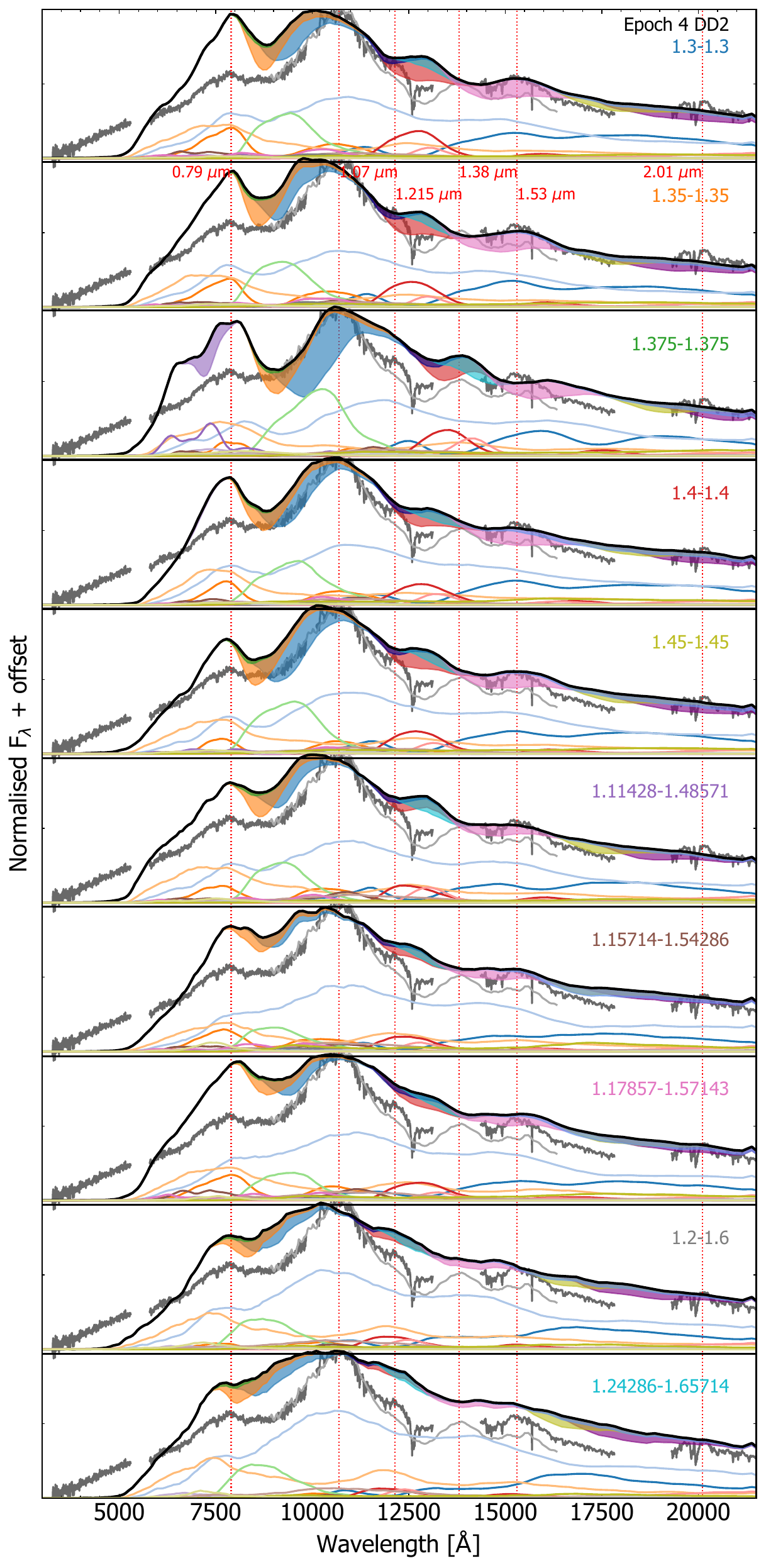}
    \caption{Same as Figure~\ref{fig:spec_emission_epoch2} for the spectra at epoch 4. Also plotted is the HST spectrum of AT2017gfo at 4.9 days \citep{tanvir2017a}.}
    \label{fig:spec_emission_epoch4}
\end{figure*}

We also show the spectra for each model in Figures~\ref{fig:spec_emission_epoch1}--\ref{fig:spec_emission_epoch4}, where each spectrum has been normalised to its peak flux (by dividing the flux at each wavelength by the maximum flux) and plotted with an offset of 1. The observed spectra of AT2017gfo are normalised in the same way and plotted for comparison to the simulated spectra.
The ion emission contributions for all models are also shown. 
In epoch 4, we additionally show the HST spectrum from \citet{tanvir2017a} at 4.9 days showing a feature at 1.4$\mu$m that is masked by telluric absorption in the ground based X-Shooter spectra \citep{pian2017a,smartt2017a}.

\subsubsection{Epoch 1}
\label{sec:epoch1}

The epoch 1 spectra are shown in Figure~\ref{fig:spec_emission_epoch1}. The ion emission contributions are shown beneath each spectrum (similarly to Figure~\ref{fig:specemissionabsorption}, except the ion contributions are not stacked).

At this epoch, the greatest contributions to the emission come from \ion{Nd}{III}, \ion{Zr}{IV} and \ion{Ce}{III}, with the exception of SFHO 1.15714--1.54286 and DD2 1.4--1.4, where \ion{Ce}{II} dominates the emission $\lesssim 9000$ \AA, indicating a lower ionisation state in these models.
SFHO 1.15714--1.54286 and DD2 1.4--1.4 have among the highest densities of Ce in the outer ejecta. 
\ion{Pr}{III}, \ion{Tb}{III} and \ion{Nd}{II} also contribute significantly at this epoch. In some models, \ion{Dy}{III} also has a relatively strong contribution around $3700$ \AA, such as in DD2 1.11428-–1.48571, DD2 1.375--1.375 and DD2 1.3--1.3. However, note that \ion{Dy}{III} is uncalibrated in our atomic data set used for the radiative transfer calculations (see discussion by \citealt{floers2026a}).
The emission from \ion{Nd}{III} and \ion{Zr}{IV} appear to shape the continuum emission $\lesssim 9000$, in addition to the \ion{Sr}{II} feature in those models where it appears.
\ion{Nd}{III} is the greatest contribution to the region around 5000 \AA, which appears as a peak in some models, even those with weaker contributions from \ion{Sr}{II}, such as SFHO 1.4--1.4, DD2 1.3--1.3 and DD2 1.11428–-1.48571.

At epoch 1, 
some of the spectra resemble the observations of AT2017gfo at wavelengths between $4000$ -- $7500$ \AA, including SFHO 1.3--1.3, SFHO 1.35--1.35 and DD2 1.45--1.45.
These models show very weak contributions from \ion{Sr}{II} at this epoch that are not distinguishable in the spectrum (see also Figure~\ref{fig:specemissionabsorption}).
This is likely because of the low density of Sr (Figure~\ref{fig:model_los}) in the polar region in these models in the outer ejecta.

As discussed above, some models already show an absorption trough between $\sim 5000$ and $7000$~\AA. 
These models show larger contributions from \ion{Sr}{II} at these wavelengths, in particular SFHO 1.15714--1.54286 and SFHO 1.2--1.6, which have the highest Sr densities in the polar region in the outer ejecta (above $\sim 0.45$c).
In all of these cases, the \ion{Sr}{II} contributions are at higher velocities (i.e. lower wavelengths) than the inferred absorption feature in AT2017gfo around 8000~\AA, attributed to \ion{Sr}{II} \citep{watson2019a}.
Most models show some emission from \ion{La}{III} $\lambda14096$, \ion{Ce}{III} $\lambda16128$ and \ion{Ce}{III} $\lambda20691$, although these are very weak at this epoch and are not yet distinguishable in the spectrum (see discussion below on the emergence of these features).

None of the models match the SED of AT2017gfo at wavelengths $\gtrsim 7500$ \AA, showing much fainter spectra at these redder wavelengths.
In the simulations, \ion{Ce}{III} is the greatest contribution to emission at these wavelengths. 
The models also do not reproduce the spectrum of AT2017gfo at wavelengths $\lesssim 4000$ \AA, showing too much flux at these wavelengths.

At this epoch, there are also contributions from other triply ionised species in addition to \ion{Zr}{IV}, including \ion{Hf}{IV} and \ion{Os}{IV} (with additional more minor contributions including \ion{Pt}{IV}, \ion{Xe}{IV} and \ion{Ce}{IV}). These species however are not calibrated to experimental values, and therefore the wavelengths and transition strengths are likely inaccurate in our simulation. To properly test the impact of these species, accurate triply ionised lanthanide atomic data are required.

\subsubsection{Epoch 2}
In addition to the ion emission contributions shown in Figure~\ref{fig:spec_emission_epoch1},  Figures~\ref{fig:spec_emission_epoch2}--\ref{fig:spec_emission_epoch4} (showing the spectra at epochs 2--4) also show the emission contributions of individual spectral lines.
These are the fraction of the ion emission contribution that are contributed by an individual spectral line of that ion.

At epoch 2 (Figure~\ref{fig:spec_emission_epoch2}),
all models now show a spectral feature from the \ion{Sr}{II} triplet, as was previously found in the simulation by \citet{shingles2023a}, with lines from \ion{Sr}{II} $\lambda10914$, $\lambda10327$ and $\lambda10036$. 
Note that while only the emission components are plotted here, these lines also show an absorption component, forming a P-Cygni like feature (see Appendix~\ref{sec:linefeatures}, Figure~\ref{fig:spec_emission_SrIILaIIIGdIII}, showing \ion{Sr}{II} absorption by these lines).

All models also show contributions from \ion{La}{III} $\lambda14096$ (in agreement with \citealt{domoto2022a}), with contributions from \ion{Gd}{III} $\lambda14332$ to the emission at the red wing of the \ion{La}{III} $\lambda14096$ emission, and \ion{La}{III} $\lambda13894$ contributes weakly to the blue side of the \ion{La}{III} $\lambda14096$.
The \ion{La}{III} lines also show an absorption component, forming a P-Cygni feature (Figure~\ref{fig:spec_emission_SrIILaIIIGdIII}). 
The \ion{Gd}{III} $\lambda14332$ line however only shows a very minor absorption component (Figure~\ref{fig:spec_emission_SrIILaIIIGdIII}).
This blended \ion{La}{III} and \ion{Gd}{III} feature was not present in the simulation by \citet{shingles2023a} because the atomic data used in their simulations for these elements was not calibrated to experimentally known energy levels \citep{tanaka2020a}. This blended feature now appears since we use the atomic data from \citet{floers2026a}.

In some models this blend of \ion{La}{III} and \ion{Gd}{III} features is clearly distinguishable in the simulated spectra, e.g., SFHO 1.35--1.35 and SFHO 1.375--1.375, while in other models the region around these features remains continuum-like.
The \ion{Gd}{III} $\lambda14332$ line was identified by \citet{gillanders2024a} and \citet{rahmouni2025a} as the strongest \ion{Gd}{III} feature that may form in kilonovae spectra, and they suggested that this would affect the \ion{La}{III} feature.
The results of our simulations agree with this.
The next strongest \ion{Gd}{III} line identified by \citet{rahmouni2025a} is $\lambda 17474$, although this contribution is only minor in our simulations and is unlikely to be observable. It would also form at the same wavelengths as the strong \ion{Ce}{III} $\lambda16128$ line in our simulations, so would be unlikely to be identifiable even if it were stronger.

The models all show contributions from \ion{Ce}{III}. The strongest lines of \ion{Ce}{III} are $\lambda16128$, $\lambda20691$ and $\lambda19524$. In some models these can be seen as clear emission features in the simulated spectra, particularly the \ion{Ce}{III} $\lambda16128$ and $\lambda20691$ lines.
At this epoch, these lines contribute between $\sim 40$--$50\%$ of the total \ion{Ce}{III} emission, with the \ion{Ce}{III} $\lambda16128$ contributing \mbox{$\sim 20$--$25\%$}, $\lambda20691$ contributing around $\sim 15\%$ and $\lambda19524$ contributing $\sim 10\%$ of the total \ion{Ce}{III} emission.
Blueward of the blended \ion{La}{III} and \ion{Gd}{III} feature, an emission contribution can be seen from \ion{Ce}{III} in most of the models which is a blend where \ion{Ce}{III} $\lambda 12756$ and $\lambda 12821$ are the strongest contributions (each typically a few percent of the total \ion{Ce}{III} flux).
\citet{domoto2022a} have previously identified \ion{Ce}{III} features in their spectra, and \citet{gillanders2024a} also identify \ion{Ce}{III} as a candidate for many features observed in AT2017gfo.

\begin{figure*}
    \centering
    \includegraphics[width=0.9\linewidth]{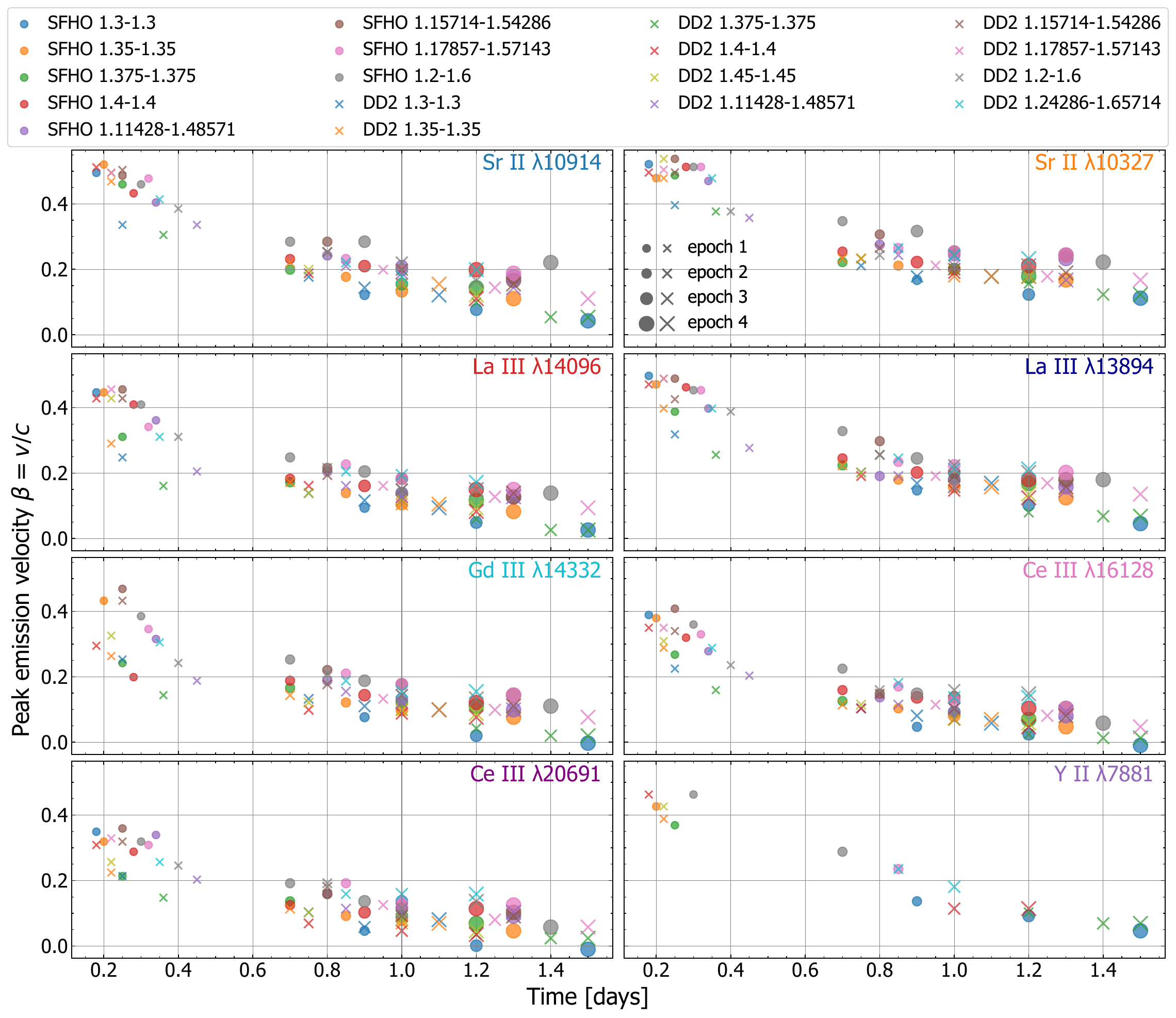}
    \caption{Blueshifted velocity of the spectral lines, measured from the peak of the emission contribution in the \textsc{artis} simulation to the rest wavelength of line for each model (i.e. blueshifts are obtained from the peak of the emission of escaping packets that were recorded as last interacting with that specific line). This gives an indication of the typical line of sight velocity shift for each line. Note that these velocity shifts are not measured from the overall spectrum, as would be typical for obtaining this information for an observed spectrum.
    }
    \label{fig:emissionvelocities}
\end{figure*}

The blueshifted velocities of these emission lines are shown in Figure~\ref{fig:emissionvelocities} for each epoch.
The velocities are obtained by taking the peak of the spectral line emission contribution (i.e. the peak of the emission of escaping photon packets where \textsc{artis} has recorded that spectral line as the last interaction of that packet) and calculating the blueshift from the rest wavelength.
As shown in Figure~\ref{fig:emissionvelocities}, the \ion{Ce}{III} $\lambda16128$ and $\lambda20691$ lines tend to form at lower velocities than the \ion{Sr}{II} lines.

DD2 1.375--1.375 shows a clear feature from \ion{Y}{II} $\lambda7881$, in contrast to the rest of the models.
The emission contributions of individual lines of \ion{Y}{II} are shown in Figure~\ref{fig:spec_emission_YII}, showing multiple emitting lines.
These emitting lines show only weak corresponding absorption, and appear to be enhanced by fluorescence, pumped by many weak lines between $~\sim 3000$ -- $5500$~\AA. 
\citet{sneppen2023d} identified that this \ion{Y}{II} $\lambda7881$ line would be the strongest \ion{Y}{II} line and identified a P-Cygni like feature in AT2017gfo that they attribute to this.
In our simulation, this feature is primarily shaped by two \ion{Y}{II} emission features, rather than by P-Cygni absorption and emission by the \ion{Y}{II} $\lambda7881$ line.
It is interesting to note that DD2 1.375--1.375 has one of the lowest densities of Y across all velocities (Figure~\ref{fig:emissionvelocities}). 
Given this, the strength of this line is likely dependent on the availability of UV photons to pump the upper energy level of the $\lambda7881$ transition, with an upper level energy of 3.4116 eV.
According to NIST, this level is \mbox{4d5p z 1P°} with a leading percentage of $50\%$, but NIST also shows this level as \mbox{5s5p 1P°} with a leading percentage of $25\%$, meaning that this level can also be pumped from the ground configuration.
The top contributing lines of \ion{Y}{II} are shown in Figure~\ref{fig:spec_emission_YII}, along with the energy level labels for each transition as given by \citet{kurucz2018a}.

At this epoch, SFHO 1.3--1.3, SFHO 1.17857--1.57143 and DD2 1.24286--1.65714 show very minor contributions from the $\lambda7881$ line, however, no models other than DD2 1.375--1.375 show a clear feature from \ion{Y}{II} $\lambda7881$ at this epoch, even though most models have a higher density of Y than DD2 1.375--1.375.
The times at which DD2 1.375--1.375 shows the closest match to the SED of AT2017gfo at each epoch are the latest out of all models, and
the spectral lines
form at relatively low velocities compared to the other models.
The low lanthanide fraction of DD2 1.375--1.375 in this line of sight is likely the reason why this model evolves differently to the other models, and likely allows the outer ejecta to be more optically thin than in other models. 
This suggests that the features form deeper in the ejecta, where there may be more blue photons pumping the upper level of the $\lambda7881$ transition, which is likely the reason the feature appears in this model.

In some of the simulations (particularly SFHO 1.15714-1.54286), a contribution from \ion{Lu}{I} $\lambda24177$ has started to appear. 
SFHO 1.15714-1.54286 showed the greatest contribution from \ion{Ce}{II} in the previous epoch, indicating a lower ionisation state. This also suggests a low ionisation state for this model.
At this time, there are minor contributions from \ion{Y}{I} and \ion{Zr}{I} (as well as \ion{Cs}{I}, however, this ion is uncalibrated in our atomic data).
This shows that even at such early times, some neutral species are able to contribute to the spectrum. However, none of these produce any clear signatures in the overall spectrum at this epoch.

At epoch 2, \ion{Ce}{II} becomes the dominant contribution to the spectra in most models, indicating that the Ce in the ejecta is less ionised than at epoch 1 where \ion{Ce}{III} generally showed a larger contribution. At wavelengths $\gtrsim 12000$ \AA \ \ion{Ce}{III} still remains an important contribution.
Note that the \ion{Ce}{III} ion contribution plotted includes the contributions from the \ion{Ce}{III} lines represented by the shading beneath the spectrum, but also many weak \ion{Ce}{III} lines emitting at these wavelengths. 
The dominance of \ion{Ce}{III} and \ion{Ce}{II} to the emission is in agreement with findings by \citet{shingles2023a}.
\ion{Zr}{IV} no longer contributes significantly at epoch 2, also showing the decrease in ionisation state compared to epoch~1.
\ion{Nd}{III} is still a significant contribution in most of the models and is still predominantly responsible for the peak that forms around $7000$ \AA \ (also at lower or higher wavelengths depending on the model), although now this region also shows significant contributions from \ion{Nd}{II}. \ion{Pr}{III} and \ion{Tb}{III} also still contribute to this region in some of the models, but no longer significantly contribute to others, e.g. SFHO 1.2--1.6, where \ion{Ce}{II} dominates across nearly all wavelengths, suggesting a lower ionisation state in this model.

The epoch 2 spectrum of AT2017gfo is also plotted in Figure~\ref{fig:spec_emission_epoch2} for comparison, and the emission-like features identified at this epoch in the observed spectrum are indicated. 
The wavelengths of the emission features marked in this plot, as well as in Figures~\ref{fig:spec_emission_epoch3} and \ref{fig:spec_emission_epoch4}, are taken from \citet{gillanders2024a}.
While most models do show similar emission-like features, they all struggle to simultaneously match the velocities of the two features (at $0.69 \mu$m and $1.015 \mu$m in the observations).

The feature at $0.69 \mu$m has previously been modelled as the blue wing of the \ion{Sr}{II} P-Cygni feature, where this joins the continuum emission \citep[e.g.,][]{watson2019a, gillanders2022a, gillanders2024a}.
\citet{sneppen2023d} suggested that the feature around $0.69 \mu$m is a P-Cygni feature from \ion{Y}{II}.
Our models show that \ion{Y}{II} can affect the spectrum in this region, such as in DD2 1.375--1.375, although the relative strength of the feature in this simulation appears stronger than in AT2017gfo.
This feature is shaped by two emission components boosted by fluorescence (as can be seen in Figure~\ref{fig:spec_emission_YII}), rather than by a P-Cygni profile alone. It is interesting to note that the remaining simulations show no clear \ion{Y}{II} feature around this wavelength, and the largest contribution to the region around $0.69 \mu$m is from \ion{Nd}{III}.

The feature indicated in AT2017gfo around $1.015 \mu$m has been identified as \ion{Sr}{II} (\citealt{watson2019a}; although see discussion of \ion{He}{I} at this wavelength: \citealt{perego2022a, tarumi2023a, sneppen2024c, arya2026a, chiba2026a}).
Our models all predict a \ion{Sr}{II} feature, although in most of the simulations the \ion{Sr}{II} forms at higher velocities than observed in AT2017gfo. In particular, the absorption component forms at higher velocities in most models, but in some models the emission peak also appears at higher velocities.
The high velocities are likely due to the earlier times in the simulations at which the SED is close to that of AT2017gfo, suggesting that the line forming region in the simulations has not receded to as low velocities as in AT2017gfo.
In many models, the emission component of the feature around $1.015 \mu$m appears to be broadened by \ion{Nd}{III} emission, peaking around this wavelength, such as SFHO~1.35--1.35 and DD2 1.45--1.45.
The velocity of the \ion{Sr}{II} feature in DD2 1.375--1.375 appears to show a good match to AT2017gfo, although the relative strength of the peak blueward of the \ion{Sr}{II} feature appears too strong.

In DD2 1.375--1.375, the blended feature of \ion{La}{III} $\lambda14096$, \ion{Gd}{III} $\lambda14332$ and \ion{La}{III} $\lambda13894$ forms in a region masked by telluric absorption in the ground based observations of AT2017gfo, although a feature appears at $\sim1.4 \mu$m in AT2017gfo in the HST spectrum taken at 4.9 days, previously suggested to be \ion{La}{III} and \ion{Gd}{III} \citep[][see Section~\ref{sec:epoch4}]{domoto2022a, gillanders2024a, rahmouni2025a}.
It is likely that this feature was present before the HST observation, but masked by telluric absorption.
As previously discussed, the spectral features in DD2 1.375--1.375 form at relatively low velocities.
In all of the simulations this blend is present, although generally it forms at higher velocities than in DD2 1.375--1.375. At this epoch this blend generally forms a detectable feature in most models.

The simulations tend to agree with the observations reasonably well at red wavelengths ($\gtrsim 12000$ \AA), however, none of the models can reproduce the flux at wavelengths $\lesssim 5000$ \AA, as all models show strong line blanketing of these wavelengths not observed in AT2017gfo.

\subsubsection{Epoch 3}
At epoch 3 (Figure~\ref{fig:spec_emission_epoch3}), generally the strengths of the features that appeared at epoch 2 have become stronger.
All models now show clear spectral features from \ion{Sr}{II}, blended \ion{La}{III} and \ion{Gd}{III} and from \ion{Ce}{III}.
The blueshifted velocities of these features generally have decreased compared to epoch 2 (see Figure~\ref{fig:emissionvelocities}), as the line forming region has receded further inwards to lower ejecta velocities.

DD2 1.375--1.375 still shows a strong \ion{Y}{II} feature and SFHO~1.3--1.3 now also shows a clear \ion{Y}{II} $\lambda 7881$ feature. Like in DD2~1.375--1.375, \ion{Y}{II} shows multiple emission peaks in SFHO~1.3--1.3, shaped by many lines of \ion{Y}{II}, which appear to be boosted by fluorescence, with many \ion{Y}{II} lines absorbing in the UV. 
Like DD2 1.375--1.375, SFHO~1.3--1.3 has a low Y density as well as a low lanthanide fraction in the polar region (although the lanthanide fraction is not as low as DD2 1.375--1.375).
In this model, spectral features also form at relatively low velocities, likely because the low lanthanide fraction in the line of sight leads to lower optical depths in the outer ejecta.
The rest of the simulations do not show an observable feature, although other models do show minor contributions from \ion{Y}{II} $\lambda 7881$.
This would suggest that if the feature at $0.75 \mu$m in AT2017gfo is \ion{Y}{II} \citep{sneppen2023d}, then this would imply that AT2017gfo also had a relatively low lanthanide fraction in the line of sight.

The \ion{Nd}{III} contributions are further reduced at epoch 3, and almost entirely gone from some models. 
\ion{Nd}{II} is now the dominant contribution to wavelengths around $7000$ \AA, although \ion{Ce}{II} dominates the continuum emission across most of the spectrum.
\ion{Ce}{III} remains important beyond $\sim 15000$ \AA.

At this epoch, additional emission-like features have emerged in the AT2017gfo spectrum, which are indicated in Figure~\ref{fig:spec_emission_epoch3}.
In some of the models, the \ion{Ce}{III} $\lambda16128$ and $\lambda20691$ features appear to align with the observed features at $1.52 \mu$m and $1.99 \mu$m, respectively, for example in DD2 1.4--1.4 and DD2 1.45--1.45. 
\ion{Ce}{III} has previously been suggested as a possible identification for these features \citep{domoto2022a, gillanders2024a}.
However, in a number of models, the velocities of these features are too high, and in some models the velocity is even too low, such as DD2~1.375--1.375.
The range in peak emission velocity of these features is shown in Figure~\ref{fig:emissionvelocities}.
Generally, the models with higher ejecta masses show spectral features at higher velocities than the models with lower ejecta masses.

In our simulations, the emission-like feature around $0.75 \mu$m is still predominantly composed of \ion{Nd}{III} and \ion{Nd}{II}, in differing ratios for different models. The \ion{Ce}{II} continuum like emission is also a large contribution at these wavelengths. \ion{Pr}{III} also contributes in some models.
The emission-like feature around $1.06 \mu$m is mostly explained by \ion{Sr}{II} emission in our simulations, but many of the models show a broadened red wing to this feature from \ion{Nd}{III} and \ion{Ce}{III} (as well as \ion{Pr}{III} in some models) in addition to the continuum-like \ion{Ce}{II} emission.
Generally the simulations show this peak at higher velocities than AT2017gfo, but in some models the velocity matches AT2017gfo, such as DD2~1.375--1.375 and DD2~1.4--1.4. The simulations still struggle to reproduce the velocity of both this emission-like peak and the peak at $0.75 \mu$m simultaneously.
As with previous epochs, wavelengths $\lesssim 5000$ \AA \ are too strongly line blanketed compared to AT2017gfo.

\subsubsection{Epoch 4}
\label{sec:epoch4}

At epoch 4 (Figure~\ref{fig:spec_emission_epoch4}), the simulated spectral features have decreased to lower velocities.
For SFHO~1.3--1.3 some lines even have a peak emission blueshift velocity of slightly less than zero.
It is likely that this is due to the redward drift described by \citet{gillanders2024a} and \citet{mcneill2026a}, where scattered photons can be redshifted due to the reverberation effect across the ejecta. 
The strong lines discussed for previous epochs continue to be present at epoch 4.
In addition to these lines, some models show a feature due to \ion{Lu}{I} $\lambda 24177$, as well as continuum-like emission from \ion{Lu}{I}, however, it is possible that this is artificially strong here due to the LTE approximation assumed in our calculations, which may underestimate the ionisation state by this epoch since non-LTE effects may prevent recombination to neutral species at such early times.

\ion{Ce}{II} dominates the continuum emission, indicating a lower ionisation state in the line forming region.
\ion{Ce}{III} remains a significant contribution at wavelengths beyond $\sim 15000$ \AA, in particular the \ion{Ce}{III} $\lambda16128$, $\lambda20691$ and $\lambda19524$ lines.

The emission-like peak around $0.79 \mu$m in the simulations is now mostly shaped by \ion{Nd}{II} and \ion{Ce}{II}, with some models still showing a contribution from \ion{Nd}{III}.
In SFHO 1.3--1.3 and DD2 1.375--1.375, a \ion{Y}{II} feature is still present around these wavelengths, to the blue side of this emission-like peak.
\ion{Sr}{II} still contributes to wavelengths around the $1.07 \mu$m emission-like peak, although at this epoch it appears the \ion{Sr}{II} emission mostly contributes to the blue side of this peak, while the red wing tends to be broadened by the \ion{Ce}{II} continuum-like emission. \ion{Nd}{II}, \ion{Nd}{III}, \ion{Ce}{III} and \ion{Pr}{III} also contribute to the red side of this emission-like peak in differing strengths across the models.
Unlike previous epochs, at this epoch some of the models form the two emission-like peaks at $0.79 \mu$m and $1.07 \mu$m, matching the velocities of these peaks in AT2017gfo, including SFHO 1.35--1.35, SFHO 1.375--1.375 and DD2~1.375--1.375. However, the relative strength of the $0.79 \mu$m peak tends to be too strong compared to the $1.07 \mu$m peak in AT2017gfo.

At this epoch, an emission-like feature at $1.215 \mu$m emerged in AT2017gfo.
In some of the simulations, this feature aligns with the blended \ion{La}{III} and \ion{Gd}{III} feature, such as the two 1.2--1.6 models, SFHO 1.15714--1.54286, SFHO 1.17857--1.57143 and DD2~1.24286--1.65714 (the models with the highest ejecta mass).
However, these are also the models where the \ion{Sr}{II} feature remains at higher velocities than AT2017gfo, suggesting that the velocity of this feature is also too high compared to AT2017gfo, and not the cause of the $1.215 \mu$m feature.
In models showing features with lower velocities, there is \ion{Ce}{III} emission around $1.215 \mu$m, or blueward of this feature. While our simulations do not produce an emission-like peak in the net spectrum, this \ion{Ce}{III} emission appears to be the most likely candidate for forming a feature in this region.
The \ion{Ce}{III} emission around $1.215 \mu$m is predominantly \ion{Ce}{III} $\lambda 12756$ and $\lambda 12821$, although there are also additional weaker lines contributing to this blend, as is shown in Figure~\ref{fig:spec_emission_CeIIINdIII_epoch4}.
In some models, \ion{Nd}{III} and \ion{Pr}{III} emission appears blueward of the $1.215 \mu$m feature.

A spectrum was taken of AT2017gfo by HST \citep{tanvir2017a} around this epoch, showing an emission feature at $1.38 \mu$m. This is shown in Figure~\ref{fig:spec_emission_epoch4}, and has been attributed to \ion{La}{III} and \ion{Gd}{III} \citep{domoto2022a, gillanders2024a, rahmouni2025a}.
A blended \ion{La}{III} and \ion{Gd}{III} feature is present in all of the simulations with varying strengths and velocities.
Interestingly the models that show a \ion{Y}{II} feature are also the models that show the best agreement in the velocity of the \ion{La}{III} and \ion{Gd}{III} feature (SFHO 1.3--1.3 and DD2 1.375--1.375). In all other models, the velocity of this feature appears too high compared to AT2017gfo.

The \ion{Lu}{I} $\lambda 24177$ feature that appears in some models, such as SFHO 1.2--1.6, does not appear to match any features in AT2017gfo, which also may suggest that the ejecta in the simulations are less highly ionised than in AT2017gfo at this epoch, since this feature does not appear to form.

The best candidates for the features at $1.53 \mu$m and $2.01 \mu$m in our simulations are the \ion{Ce}{III} $\lambda 16128$ and $\lambda 20691$ lines, although in some models these form at too high or too low velocity compared to AT2017gfo.

\section{Discussion and conclusions}

We have calculated kilonova light curves and spectra from a range of dynamical ejecta models, simulated from hydrodynamical neutron star merger simulations \citep{vijayanthesis} with nuclear network calculations.
We adopt the same methodology as \citet{shingles2023a}, who considered only a single merger model. Here we consider a range of neutron star masses and two different equations of state.
In the radiative transfer simulations, we have included new calibrated atomic data for singly and doubly ionised lanthanides \citep{floers2026a}, allowing lanthanide spectral features to be predicted which were not present in the simulations of \citet{shingles2023a} using the uncalibrated lanthanide data from \citet{tanaka2020a}.

We presented bolometric light curves and spectra in the direction of the poles, near the observed viewing direction of AT2017gfo.
The simulated light curves are fainter and fade faster than was observed for AT2017gfo, likely due to the low ejecta mass in the models resulting from including only dynamical ejecta.
Models with higher ejecta masses generally show brighter light curves and slower decline rates than models with lower ejecta masses. 

Despite the low ejecta masses,
the simulated spectra go through phases where they resemble the observed spectral series of AT2017gfo.
We have compared the simulated spectra to AT2017gfo at times where the simulated SED shows the closest match to the observed SED.
These times, however, are much earlier than the observed times of AT2017gfo.

All the models show a similar overall evolution, with an SED initially peaking at blue wavelengths and evolving to redder colours with time.
The models share the same strong features. These include a \ion{Sr}{II} P-Cygni feature from the triplet $\lambda 10914$, $\lambda 10327$ and $\lambda 10036$ lines, a blended \ion{La}{III} and \ion{Gd}{III} feature from \ion{La}{III} $\lambda 14096$, $\lambda 13894$ and \ion{Gd}{III} $\lambda 14332$, as well as \ion{Ce}{III} features from $\lambda 16128$, $\lambda 20691$ and $\lambda 19524$.

We show that these simulated features appear to correspond with identified emission-like features in AT2017gfo. In the epoch 4 spectrum of AT2017gfo, these features form at $0.79 \mu$m, $1.07 \mu$m, $1.215 \mu$m, $1.38 \mu$m, $1.53 \mu$m and $2.01 \mu$m.
We find that in our simulations, similar features to these form, although often at bluer wavelengths. Our simulations suggest that the $1.07 \mu$m feature is from the \ion{Sr}{II} triplet, the $1.38 \mu$m feature is blended \ion{La}{III} and \ion{Gd}{III}, the $1.53 \mu$m feature can be explained by \ion{Ce}{III} $\lambda 16128$ and the $2.01 \mu$m is best described by \ion{Ce}{III} $\lambda 20691$ and $\lambda 19524$.
The best candidate in our simulations to match the $1.215 \mu$m feature is also \ion{Ce}{III} emission.
The emission-like feature at $0.79 \mu$m may be partly shaped by \ion{Y}{II} fluorescent emission, of which $\lambda 7881$ is the strongest line.
However, only two of the models show a \ion{Y}{II} feature and in most models this region is primarily shaped by \ion{Nd}{III} or \ion{Nd}{II} and \ion{Ce}{II} at the blue wing of the \ion{Sr}{II} P-Cygni feature.

The two models that show a clear \ion{Y}{II} feature are DD2 1.375--1.375 and SFHO 1.3--1.3. These models do not have particularly high Y abundances, but they do have low lanthanide densities, suggesting that the outer ejecta may be less optically thick in these models, allowing spectral features to form at lower velocities.
It may be that at these lower velocities, the radiation field is bluer, and more UV photons are available to be absorbed and fluoresce as \ion{Y}{II} emission features.
Fluorescence appears to be an important process in our kilonova simulations, in particular for forming the \ion{Y}{II} feature.

We find that \ion{Ce}{III} and \ion{Ce}{II} show important contributions to shaping the continuum, as has previously been discussed by \citet{shingles2023a}.
We also find that at epoch 1 \ion{Nd}{III} and \ion{Zr}{IV} show large contributions.
By epoch 2, \ion{Zr}{IV} no longer significantly contributes, but \ion{Nd}{III} remains important, as well as \ion{Nd}{II}, \ion{Pr}{III} and \ion{Tb}{III}.
These remain the greatest contributions until epoch 4, where \ion{Ce}{II} and \ion{Nd}{II} tend to dominate across all wavelengths, indicating that singly ionised is likely the dominant ionisation stage by this epoch.
Again, the ionisation state may be artificially low in our simulations due to the LTE approximation.

The improved atomic data for singly and doubly ionised lanthanides from \citet{floers2026a} has enabled lanthanide features to be predicted in our simulations that are in good agreement with observed features of AT2017gfo.
In particular, this new data has allowed the blended \ion{La}{III} and \ion{Gd}{III} feature to form, as well as individual \ion{Ce}{III} features.
During the times at which the simulated spectra resemble the observations, both triply ionised and neutral species contribute to the spectra. Triply ionised species, in particular, are present at the first epoch. Neutral species primarily contribute at epoch 4, but there are minor contributions from \ion{Lu}{I} as early as epoch 2.
We therefore require additional atomic data with calibrated energy levels for neutral and triply ionised species (in addition to the singly and doubly ionised data from \citealt{floers2026a}) to accurately assess if observable features could form for these ionisation states. 

We note that we do not find as significant a difference when comparing models that use the atomic data from \citet{floers2026a} and those that do not (i.e. the models of \citealt{shingles2023a}) as reported by \citet{gillanders2026a}.
In the simulation by \citet{shingles2023a}, the singly and doubly ionised lanthanide data was sourced from \citet{tanaka2020a}, which already contained many lines and therefore the lanthanides already contributed significant opacity to the simulations by \citet{shingles2023a}.
When using the lanthanide data from \citet{floers2026a}, \citet{gillanders2026a} find that their models require a lanthanide fraction of $2.5 \times 10^{-3}$. 
While we do not find perfect agreement between our models and the observations of AT2017gfo, the lanthanide fractions in our models appear to be compatible with the observations. In the line of sight, the highest lanthanide fraction in our models is $1.7 \times 10^{-2}$ (the mass weighted mean lanthanide fraction), and the highest spherically averaged mass weighted mean lanthanide fraction is $3.3 \times 10^{-2}$.

The simulations deviate from the observed spectra of AT2017gfo at the bluest wavelengths ($\lesssim 4000$ \AA). The models tend to show too strong line blanketing at these wavelengths compared to the observations.
The blue end of the spectrum is best matched at epoch 1, but by epoch 2 all models are much too faint in this region.
However, at epoch 1, all of the models are too faint at wavelengths $\gtrsim 7500$ \AA \ compared to AT2017gfo.
By epoch 2 the simulations show better agreement at these wavelengths.
The better agreement of the blue end of the spectrum at epoch 1 may be because the ionisation state is highest at epoch 1 but becomes too low by epoch 2, which may be a result of the LTE approximation.
Non-LTE simulations may lead to a higher ionisation state which could reduce the level of line blanketing, and may also reduce the appearance of neutral spectral features, particularly at epoch 4. 
\citet{brethauer2025a} have shown that accounting for non-thermal ionisation in kilonovae leads to a higher ionisation state and reduced line blanketing.

We have shown that dynamical ejecta from a range of different binary neutron star merger simulations with different binary configurations can produce spectral features that resemble those observed in AT2017gfo. However, as previously found by \citet{shingles2023a}, the times at which the simulations resemble the spectra are too early.
Across all of the models, we find that the same strong spectral features persist, including the \ion{Sr}{II} triplet, a blended \ion{La}{III} and \ion{Gd}{III} feature and \ion{Ce}{III} features. The exception to this is the \ion{Y}{II} feature, which only appears in two of the models.
This similarity in the spectra for a relatively broad range of parameters suggests that the kilonova arising from the dynamical ejecta component in the polar region could be a rather robust prediction.
This could suggest that future observations of kilonovae in this direction may also show the same strong spectral features as were observed in AT2017gfo, and the early spectral evolution of future kilonovae may appear similar.
In the future, the impact of later ejecta components on these spectral predictions should be tested.

\begin{acknowledgements}
CEC is funded by the European Union’s Horizon Europe
research and innovation programme under the Marie Skłodowska-Curie grant
agreement No.~101152610.
The authors acknowledge support of the European Research Council (ERC) under the European Union's Horizon 2020 research and innovation programme (KILONOVA No.~885281, HeavyMetal No.~101071865), by the Deutsche Forschungsgemeinschaft (DFG, German Research Foundation) through Project - ID 279384907 – SFB 1245 and MA 4248/3-1, and by the State of Hesse within the Cluster Project ELEMENTS. ZX acknowledges support by the ERC through grant NeuTrAE (No.~101165138).
We acknowledge EuroHPC JU for awarding the project ID EHPC-REG-2025R01-137 access to
resources on MeluXina hosted by LuxProvide, Luxembourg.
This research was supported in part by the cluster computing resource provided by the IT Department at the GSI Helmholtzzentrum für Schwerionenforschung, Darmstadt, Germany.
This work used the DiRAC Memory Intensive service Cosma8 at Durham University, managed by the Institute for Computational Cosmology on behalf of the STFC DiRAC HPC Facility (www.dirac.ac.uk). The DiRAC service at Durham was funded by BEIS, UKRI and STFC capital funding, Durham University and STFC operations grants. DiRAC is part of the UKRI Digital Research Infrastructure.
\href{https://github.com/artis-mcrt/artis}{\textsc{artis}}\footnote{\href{https://github.com/artis-mcrt/artis/}{https://github.com/artis-mcrt/artis/}}
\citep{artis2025a} was used to carry out the radiative transfer simulations.
\href{https://github.com/artis-mcrt/artistools}{\textsc{artistools}}\footnote{\href{https://github.com/artis-mcrt/artistools/}{https://github.com/artis-mcrt/artistools/}}
\citep{Shingles_artistools_2026}
were used for data processing and plotting.
\textsc{artisatomic} was used to generate atomic data files for \textsc{artis}. The data from Kurucz were extracted using the \textsc{carsus} package.
\end{acknowledgements}



\bibliographystyle{aa}
\bibliography{astrofritz} % if your bibtex file is called example.bib

%\begin{thebibliography}{99}
% \end{thebibliography}

%%%%%%%%%%%%%%%%%%%%%%%%%%%%%%%%%%%%%%%%%%%%%%%%%%

%%%%%%%%%%%%%%%%% APPENDICES %%%%%%%%%%%%%%%%%%%%%

\begin{appendix}
\nolinenumbers

\section{Additional nucleosynthesis information}
\label{appendix}

\begin{figure}
    \centering
    \includegraphics[width=0.9\linewidth]{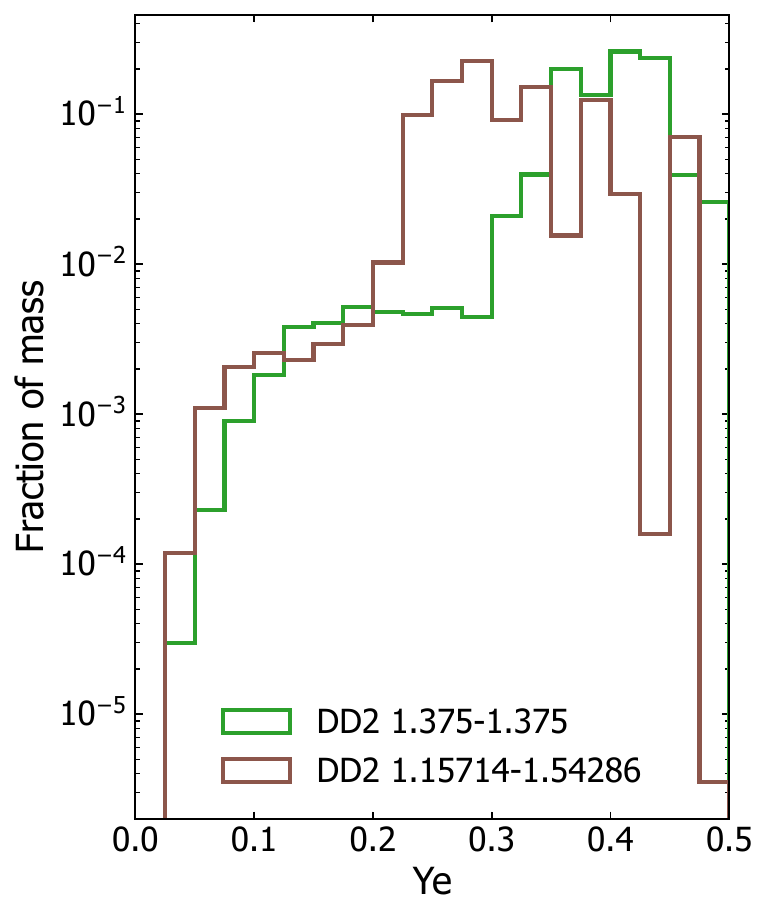}
    \caption{Fraction of the total mass ejected in the direction of the pole (given by all model grid cells whose mid-points lie within the region $<37^\circ$ from the pole, binned by the mean $Y_\mathrm{e}$ in each model grid cell (where $\Delta$$Y_\mathrm{e}$=0.025).
    The electron fraction $Y_\mathrm{e}$ is evaluated at 5~GK.
    This is shown for DD2 1.15714--1.54286 (which has the highest lanthanide fraction) and DD2 1.375--1.375 (with the lowest lanthanide fraction).
    }
    \label{fig:Ye_distribution}
\end{figure}

We show the distribution of $Y_\mathrm{e}$ across the polar region ($< 37^\circ$) in Figure~\ref{fig:Ye_distribution} for two models: DD2 1.15714--1.54286 (which has the highest lanthanide fraction) and DD2 1.375--1.375 (with the lowest lanthanide fraction).

\begin{figure*}
    \centering
    \includegraphics[width=0.9\linewidth]{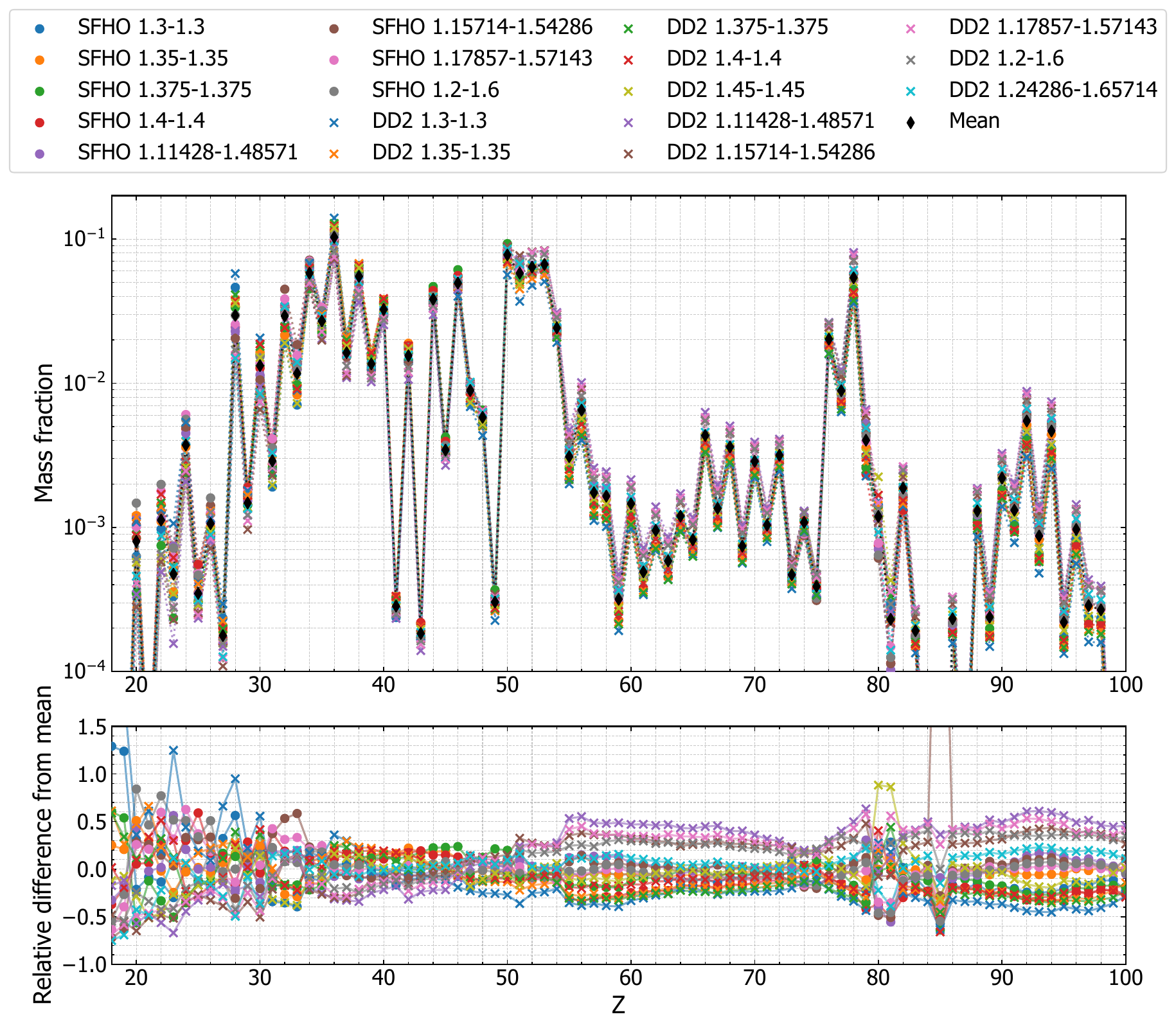}
    \caption{Upper panel: mass fractions of elements across the entire 3D ejecta. Circle markers show models with the SFHO equation of state while the x markers show those with the DD2 equation of state. Black diamonds show the mean value across all models. The lower panel shows the relative difference of each model from this mean value.}
    \label{fig:mass_fractions_entire_ejecta}
\end{figure*}

To provide additional context on the abundance patterns in the 3D dynamical ejecta models, we show the mean mass fractions across the entire ejecta in Figure~\ref{fig:mass_fractions_entire_ejecta}.
The abundance pattern in the models appears quite robust, with only small differences between models.

\section{Individual spectral line contributions}
\label{sec:linefeatures}

\begin{figure}
    \centering
    \includegraphics[width=0.9\linewidth]{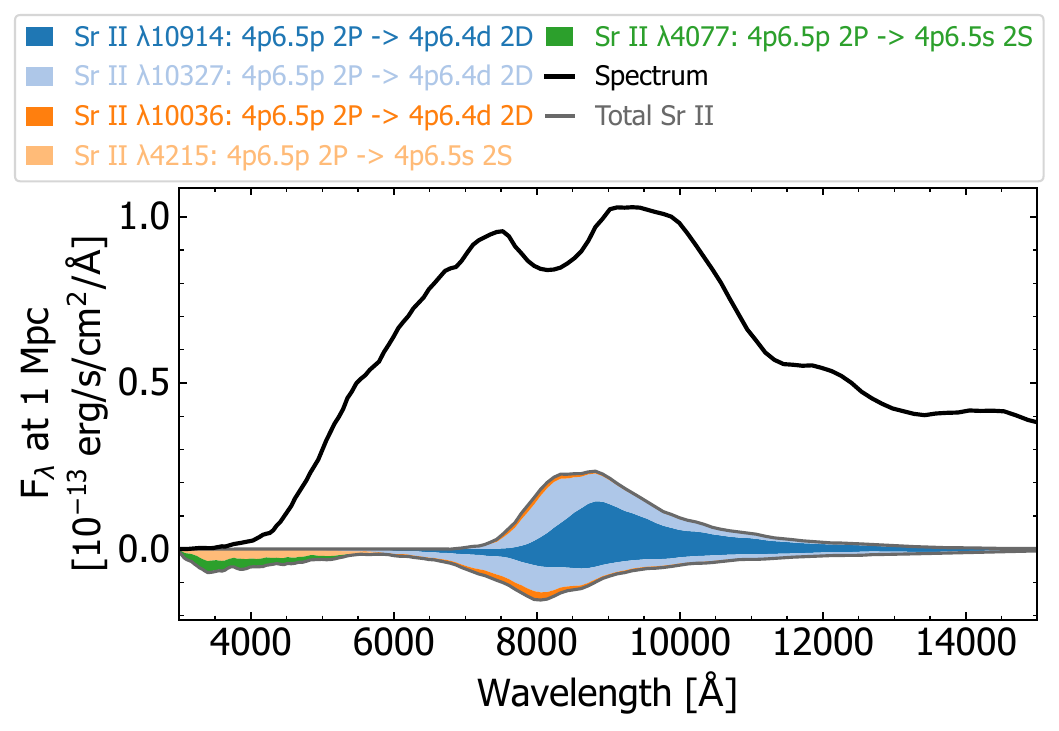}
    \includegraphics[width=0.9\linewidth]{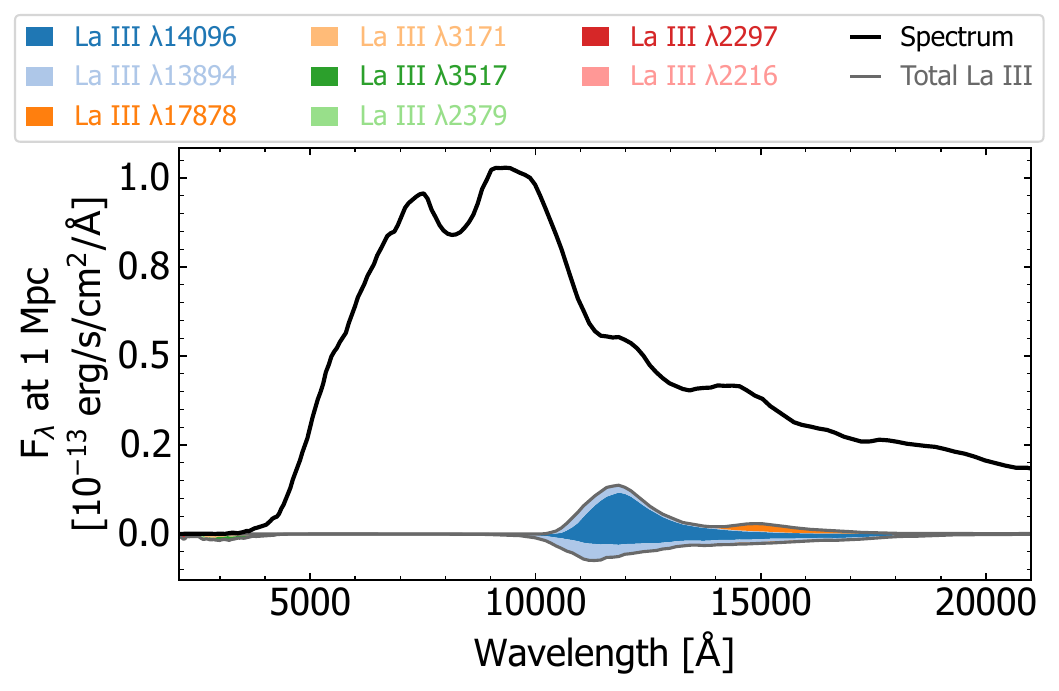}
    \includegraphics[width=0.9\linewidth]{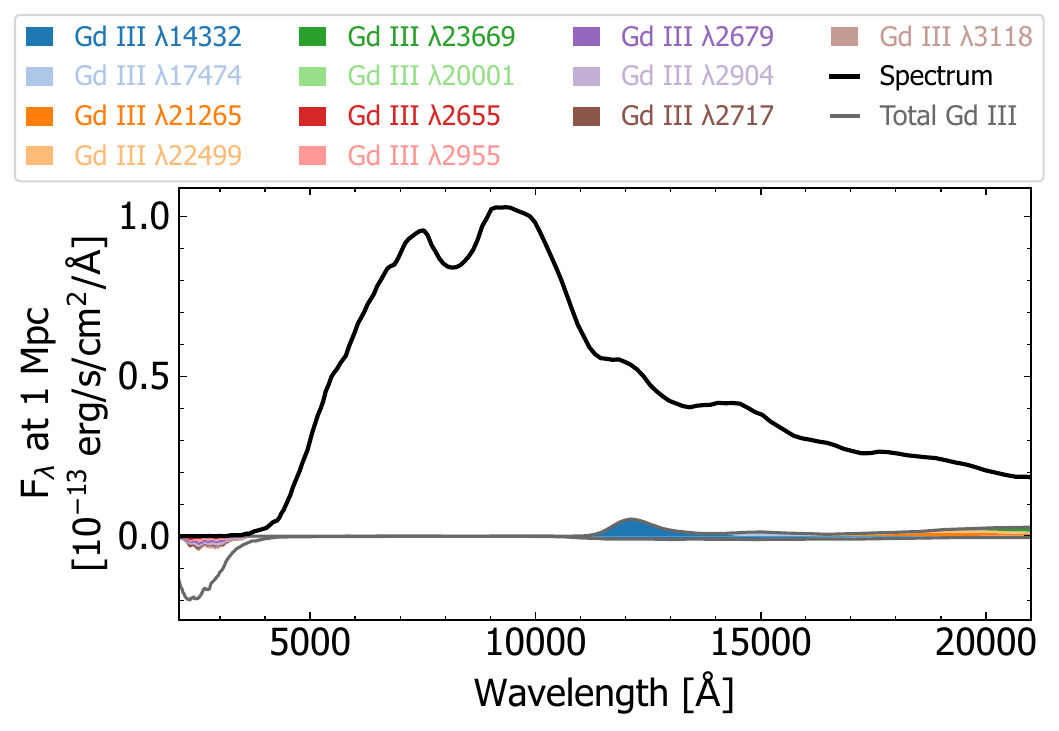}
    \caption{Spectra of SFHO 1.375-1.375 at epoch 2 showing the main individual spectral lines of an ion that contribute to the spectrum, where the grey lines show the total contribution of that ion. The upper panel shows lines of \ion{Sr}{II}, the middle panel shows \ion{La}{III} and the lower panel shows \ion{Gd}{III}. The wavelength of each line transition is given in the caption. For \ion{Sr}{II}, the energy levels of the transition are also given.
    Note that for \ion{Gd}{III} many small line contributions form the total absorption around $\sim 2500$~\AA.}
    \label{fig:spec_emission_SrIILaIIIGdIII}
\end{figure}

\begin{figure}
    \centering
    \includegraphics[width=0.9\linewidth]{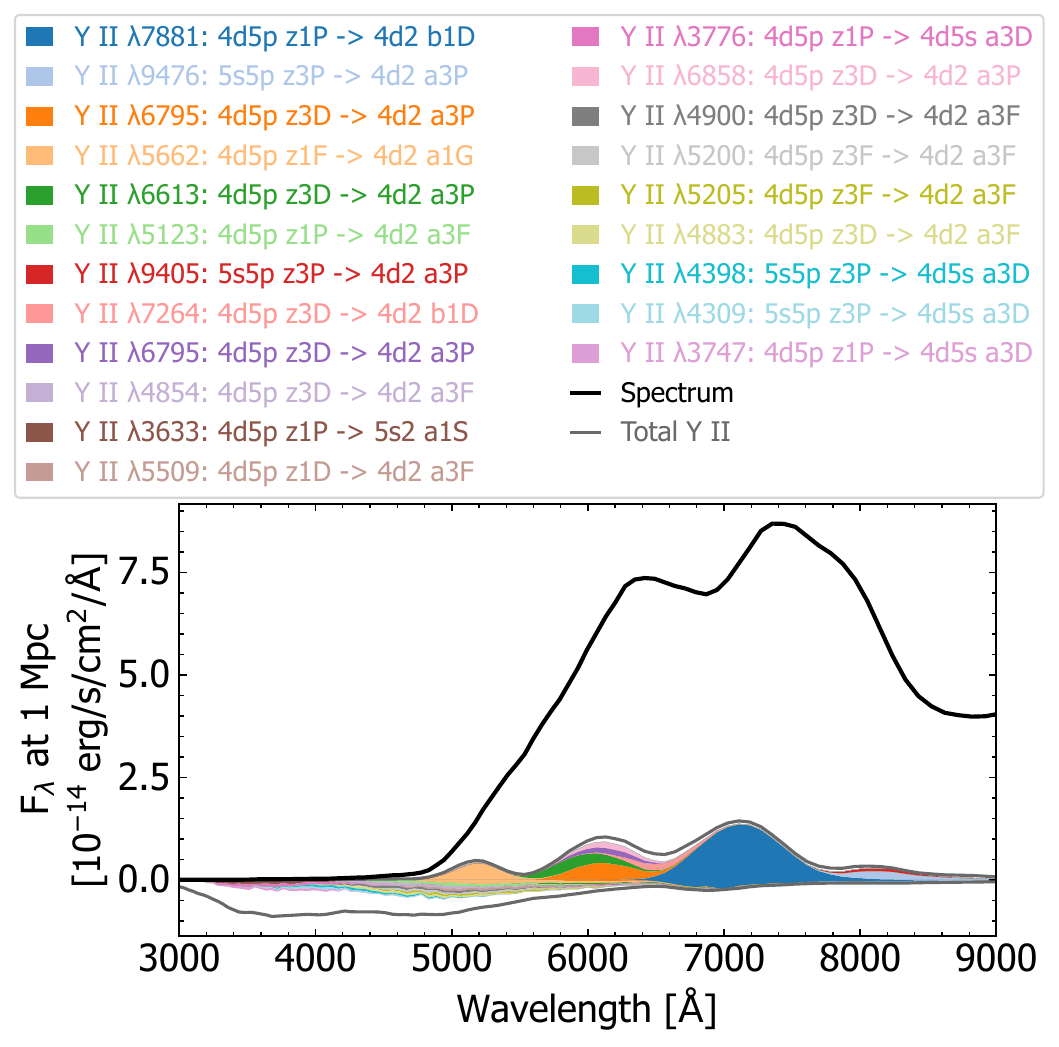}
    \caption{Same as Figure~\ref{fig:spec_emission_SrIILaIIIGdIII} but for \ion{Y}{II}. The spectrum of DD2~1.375--1.375 at epoch 2 is plotted and the top contributing lines of \ion{Y}{II} are shown. The total emission and absorption by \ion{Y}{II} are also plotted. There are many small absorption line contributions from \ion{Y}{II} lines that form the total absorption.}
    \label{fig:spec_emission_YII}
\end{figure}

\begin{figure}
    \centering
    \includegraphics[width=0.9\linewidth]{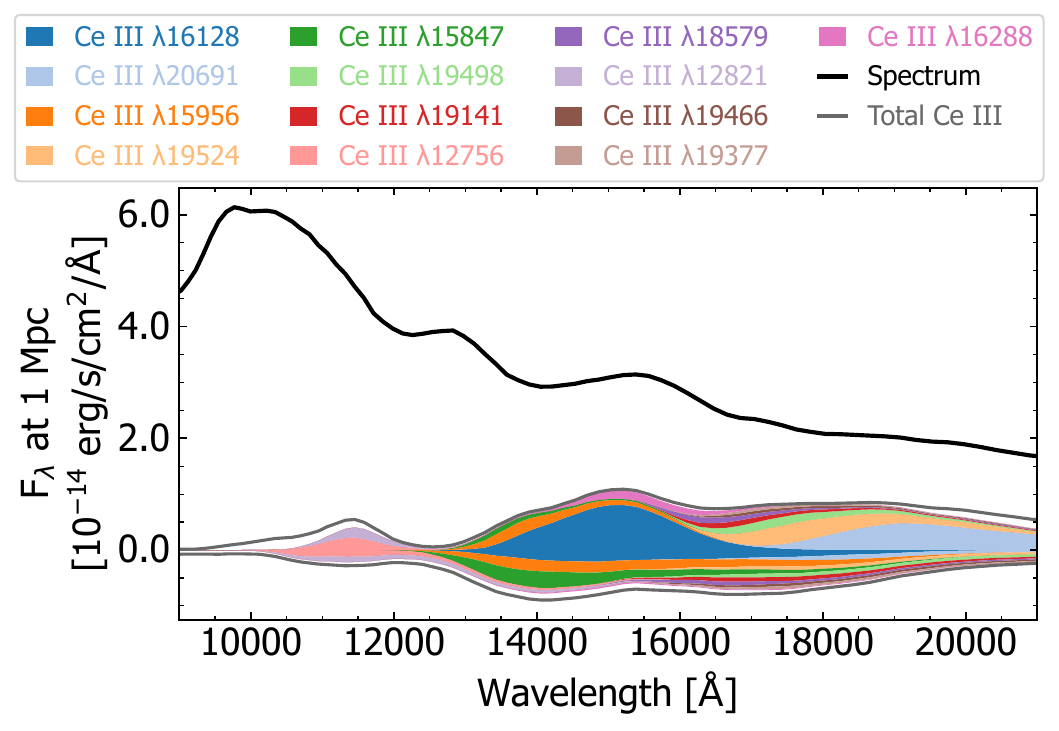}
    \includegraphics[width=0.9\linewidth]{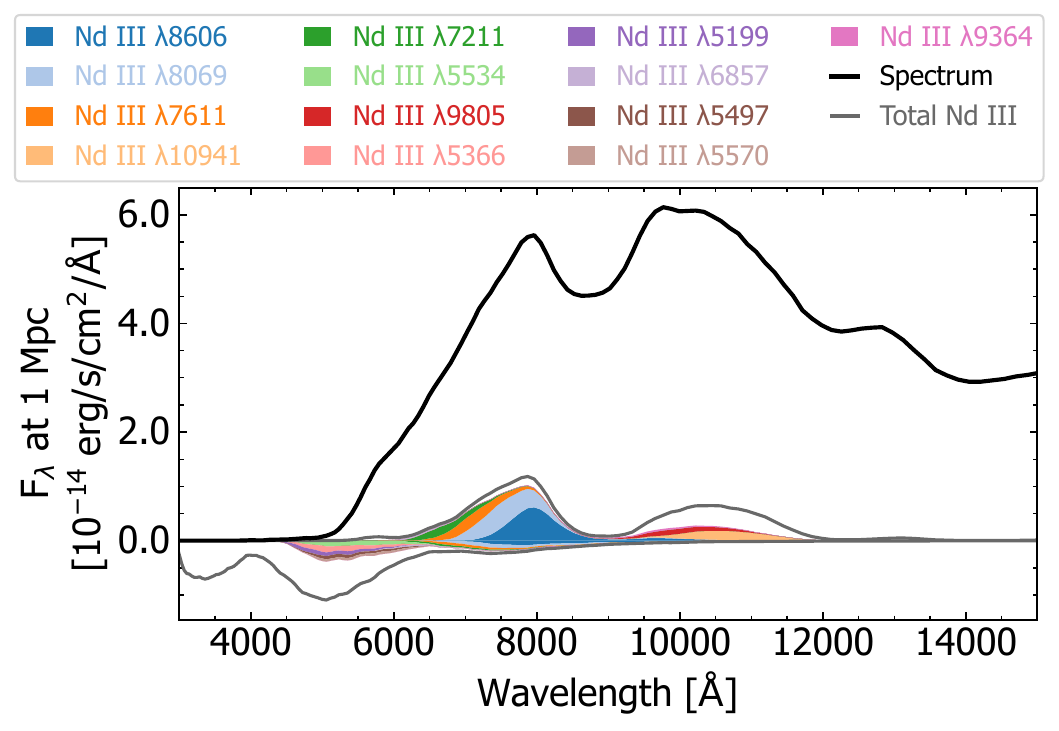}
    \caption{Same as Figure~\ref{fig:spec_emission_SrIILaIIIGdIII} but for DD2 1.35-1.35 at epoch 4 and showing the main spectral lines of \ion{Ce}{III} (upper panel) and \ion{Nd}{III} (lower panel).
    }
    \label{fig:spec_emission_CeIIINdIII_epoch4}
\end{figure}

In Figures~\ref{fig:spec_emission_SrIILaIIIGdIII}--\ref{fig:spec_emission_CeIIINdIII_epoch4}, we show the main individual line contributions responsible for the overall ion contribution in the simulations.
 Emission from individual lines is plotted above the F$_\lambda$=0 axis, 
where the shaded area represents the energy emitted by a given line. Beneath the axis, the lines that were last responsible for absorbing a photon (as recorded by \textsc{artis}) before it went on to escape the ejecta are plotted, 
indicating the amount of flux absorbed by key spectral lines.

\end{appendix}

% \appendix

% \section{Some extra material}

% If you want to present additional material which would interrupt the flow of the main paper,
% it can be placed in an Appendix which appears after the list of references.

%%%%%%%%%%%%%%%%%%%%%%%%%%%%%%%%%%%%%%%%%%%%%%%%%%

% Don't change these lines
% \bsp	% typesetting comment
\label{lastpage}
\end{document}